\documentclass[showpacs,preprintnumbers,amsmath,amssymb,12pt,floatfix]{revtex4}
\usepackage[pdftex]{graphicx}
\usepackage[usenames, dvipsnames]{color}
\usepackage{amssymb}
\usepackage{bm}
\usepackage{graphicx}
\usepackage{revsymb}
\usepackage{amsmath}
\usepackage{dcolumn}
\usepackage{ulem}

\usepackage{color} % to paint the words to be indexed

\newcommand{\be}{\begin{eqnarray}}
\newcommand{\ee}{\end{eqnarray}}

\begin{document}
\draft
\title{Gamow-Teller strength of $^{12,14,16}$C within deformed quasiparticle random-phase approximation}

\author{Eunja Ha \footnote{ejaha@hanyang.ac.kr}}
\address{Department of Physics and Research Institute for Natural Science, Hanyang University, Seoul, 04763, Korea}
\author{Myung-Ki Cheoun \footnote{Corresponding Author: cheoun@ssu.ac.kr}}
\address{Department of Physics and Origin of Matter and Evolution of Galaxies (OMEG) Institute, Soongsil University, Seoul 156-743, Korea}
\author{H. Sagawa \footnote{sagawa@ribf.riken.jp}}
\address{RIKEN, Nishina Center for Accelerator-Based Science, Wako 351-0198, Japan} 
\address{Center for Mathematics and Physics, University of Aizu, Aizu-Wakamatsu, Fukushima 965-8560, Japan}
\address{Institute of Theoretical Physics, Chinese Academy of Sciences, Beijing 100190, China}
\author{Gianluca Col\`o \footnote{colo@mi.infn.it}}
\address{Dipartimento di Fisica, Universit\`a degli Studi and INFN via Celoria 16, 20133 Milano, Italy}

\begin{abstract}
We investigate the Gamow-Teller (GT) transition strength distributions in the light carbon isotopes $^{12,14,16}$C within the framework of the deformed quasiparticle random-phase approximation (DQRPA). Nuclear deformation is explicitly incorporated through Skyrme Hartree-Fock mean-field calculations combined with the QRPA formalism. The residual particle-hole $(p-h)$ and particle-particle $(p-p)$ interactions are derived from Br\"uckner $G$-matrix calculations based on the CD-Bonn potential, and their impact on the low-lying GT strengths is systematically examined by varying the corresponding interaction strengths.

We find that nuclear deformation, associated with a reduced spin-orbit strength, plays a significant role in interpreting the GT strength distribution of $^{12}$C. In contrast, the calculated GT$^{(-)}$ strength distribution of $^{14}$C in the spherical limit reproduces the essential features of the experimental $(p,n)$ charge-exchange data. The case of $^{16}$C reveals additional high-lying GT strength associated with deformation-induced configuration mixing.

\end{abstract}

\date{\today}

\maketitle

\section{Introduction}

The Gamow-Teller (GT) excitation is one of the key transitions {for the study of} {the weak interaction of nuclei through beta decays, electron or muon capture reactions and, indirectly yet effectively, as discussed below, through charge-exchange (CEX) reactions via strong interaction.} Recent remarkable progress in the study of supernova (SN) neutrinos has revealed that the GT transition is the dominant transition in the neutrino-induced reactions relevant to the neutrino processes   \cite{Taka2015,Bala2015,Taka2013,Cheoun2012,Moha2023,Suzuki2023,Cheoun2023}. For example, the GT transition turns out to be dominant among various multipole transitions in the neutrino-nucleus ($\nu$-A) reaction, according to recent theoretical calculations \cite{Taka2015,Bala2015}. However, sufficient experimental data are still lacking to confirm the reliability of  theoretical models. Only limited data for the neutrino-induced reactions are available until now, despite many discussions about the possibility of low-energy neutrino sources \cite{Sato2005,Shin2016}. {It is expected that the recent neutrino reaction data in JSNS$^2$ experiments \cite{JSNS2,JSNS22025} and the LSND data \cite{LSND1,LSND2} could provide valuable information on low-energy neutrino reactions on various nuclear targets.}

Recent  CEX reaction data by proton or neutron beams at RIKEN or NSCL \cite{Sasano2011,Sasano2012}, triton and {$^{3}$He} beam at RCNP \cite{Freke2016} are of great help for understanding the neutrino-induced reactions because the main contribution to the neutrino-induced reaction, namely, the GT transition, can be studied experimentally by the CEX reaction. {These are governed by the strong interaction, but well-established calibration procedures allow extracting the GT matrix elements from them, so that one can use the same matrix elements to estimate the weak interaction processes.}

From the nuclear physics viewpoint, the GT transition is an allowed spin-isospin excitation of a nucleus, and the transition operator is one of the simplest. Consequently, it may give invaluable information {on the spin-isospin channel of the nuclear Hamiltonian and the spin-isospin correlations.} However, there still remain {several open problems as discussed below}, that prevent us from having a full quantitative understanding of GT excitations throughout the nuclear chart. 

In this work, we focus on one of the open questions, {deformation effect} on {GT strength in light nuclei.} {To this end, we adopt a  deformed quasi-particle random phase approximation  (DQRPA) model}. Many nuclei in the nuclear chart are considered to be  deformed, which can be verified from the E2 transition data as well as the rotational band structure. Moreover, the deformation may produce unusual phenomena of shell evolution of protons and neutrons {\cite{Ichik2019}}, which make new sub-magic numbers and change to some extent the nature of  particle-particle ($p-p$) and  particle-hole ($p-h$) interactions. One of the practical but successful ways for properly describing the deformation is to start from the Nilsson model with the axial symmetry. Our previous calculations \cite{Ha2015a,Ha2015b,Ha2016} for the GT strength distribution of open shell nuclei by deformed QRPA showed that the deformation markedly alters the GT strength peaks {compared with those }obtained by a spherical QRPA.

{For light $p$-shell nuclei such as carbon, where the spin-orbit interaction is relatively moderate, the intermediate coupling scheme between the $jj$- and  $LS$-coupling schemes is  justified \cite{Cohen-Kurath}. 
In this regime, the total spin $J$ remains a good quantum number. As a result, GT transitions, driven purely by spin-isospin (${\bm \sigma}\cdot{\bm \tau}$) operators, can be described predominantly as coherent $J=1$ excitations rather than as fragmented single-particle transitions characteristic of the strong $jj$-coupling limit. Consequently, the intermediate coupling scheme  provides a natural and efficient description of the dominant GT strength in carbon isotopes, consistent with the observed concentration of GT strength at low excitation energies and {almost all consumption of}  the Ikeda sum rule. Here we note that the reduction of the spin-orbit coupling strength may give rise to the deformation of $^{12}$C \cite{Sagawa04}.}

The primary aim of the present work is to perform DQRPA calculations with a realistic residual interaction {determined from  the Br\"uckner $G$-matrix with} the CD-Bonn potential, {and apply to study the GT strength in C isotopes.} {We utilized as a starting point the Skyrme-type mean field (MF) \cite{Stoitsov} as well as the Woods-Saxon mean field employed in the previous calculations \cite{Ha2015a,Ha2015b,Ha2016,Ch93,pan}.} Therefore, this work is an extension of our recent works \cite{Ha2015a,Ha2015b} for the DQRPA, in which all effects of the deformation are consistently treated in the QRPA framework based on the Skyrme-type mean field \cite{Ha2024}.

As an application of the present DQRPA model, we choose the GT transitions of three C isotopes, $^{12,14,16}$C. The nucleus $^{12}$C is central to nucleosynthesis, formed through the triple-alpha process {thanks to} the Hoyle state at 7.65 MeV. It further contributes to stellar evolution via $^{12}$C + $^{12}$C fusion {\cite{21,22,23} and the CNO cycle influencing the production of heavier elements and chemical enrichment of the universe. From the weak-interaction perspective, GT transitions in $^{12}$C are essential, because reactions such as $^{12}$C($\nu_{e}, e^{-}) ^{12}$N$^*$ and  $^{12}$C($\nu_{\mu}, \mu^{-}) ^{12}$N$^*$ serve as key channels for supernova neutrino study. Furthermore, their reactions become recently available using the neutrino beams from decay at rest of pions as well as kaons \cite{MiniBooNE,JSNS2}. Accurate modelling therefore requires reliable GT strength distributions and the inclusion of {possible deformation effects}. The nearby isotopes $^{14}$C and $^{16}$C provide complementary insights: $^{14}$C's long half-life underpins radiocarbon dating and stellar processes, while the quenched B(E2) in $^{16}$C \cite{Sagawa04} {may indicate some exotic feature} affecting weak-interaction rates. Thus, carbon isotopes offer critical benchmarks for studying deformation, shell evolution, and residual interactions in GT transitions relevant to {astrophysical applications}. 

Our paper is organized as follows. In Sec. II, we briefly outline the formalism, emphasizing the treatment of deformation in the DSHFB+QRPA approach with residual interactions. In Sec. III, we present numerical results for the GT strength distributions of the carbon isotopes and discuss their sensitivity to deformation and residual interactions. Finally, conclusions and perspectives for future studies are summarized in Sec. IV.

\section{formalism}

Our calculations are carried out within the QRPA framework \cite{Ha2015a}, which adopts the SHF result \cite{Stoitsov} for the mean field, and the residual interactions calculated by the Br\"uckner $G$-matrix based on the CD-Bonn potential.  On top of the mean field, the pairing correlations are taken into account in the BCS approximation. Hereafter, we briefly summarize the DQRPA model, which will be applied for calculations of the GT strength distributions. We adopt the standard QRPA formalism based on the equation of motion for the following phonon operator, acting on the BCS ground state \cite{Ha2015a}: 
\begin{equation}\label{phonon}
{\cal Q}^{\dagger}_{m,K}  =\sum_{\rho_{\alpha} \alpha \alpha'' \rho_{\beta} \beta \beta''}
[ X^{m}_{( \alpha \alpha'' \beta \beta'')K} A^{\dagger}( \alpha \alpha'' \beta \beta'' K)
- Y^{m}_{( \alpha \alpha'' \beta \beta'')K} {\tilde A}( \alpha \alpha'' \beta \beta'' K)]~,
\end{equation}
{where $\rho_{\alpha (\beta)} (= \pm 1)$ denotes the sign of the total angular momentum projection of the $\alpha$ state for the reflection symmetry.}
Here, we have introduced pair creation and annihilation operators, composed by two quasiparticles and defined as
\begin{equation}
 A^{\dagger}( \alpha \alpha'' \beta \beta'' K)  =
 {[a^{\dagger}_{ \alpha \alpha''} a^{\dagger}_{\beta \beta''}]}^K,
~~~{\tilde A}( \alpha \alpha'' \beta \beta'' K)  =
 {[a_{\beta \beta''} a_{\alpha \alpha''}]}^K,
\end{equation}
where $K$ is the quantum number associated with the projection of the intrinsic angular momentum on the symmetry axis, which is a good quantum number in the axially deformed nuclei.
We note that parity is also treated as a good quantum number in the present approach.  
Here, $\alpha$ indicates a set of quantum numbers to specify the single-particle-state (SPS). Isospin of the real particle is denoted by the Greek letter with prime $(\alpha' , \beta' , \gamma' , \delta')$ (see Eqs. (\ref{eq:mat_A}) and (\ref{eq:mat_B})). Within the quasi-boson approximation for the phonon operator, we obtain the following QRPA equation for describing the correlated DQRPA ground state:

\begin{eqnarray}\label{qrpaeq}
&&\left(\begin{array}{cccccccc}
           A_{\alpha \beta \gamma \delta (K)}^{1111} & A_{\alpha \beta \gamma \delta (K)}^{1122} &
           A_{\alpha \beta \gamma \delta (K)}^{1112} & A_{\alpha \beta \gamma \delta (K)}^{1121} &
           B_{\alpha \beta \gamma \delta (K)}^{1111} & B_{\alpha \beta \gamma \delta (K)}^{1122} &
           B_{\alpha \beta \gamma \delta (K)}^{1112} & B_{\alpha \beta \gamma \delta (K)}^{1121} \\
           A_{\alpha \beta \gamma \delta (K)}^{2211} & A_{\alpha \beta \gamma \delta (K)}^{2222} &
           A_{\alpha \beta \gamma \delta (K)}^{2212} & A_{\alpha \beta \gamma \delta (K)}^{2221} &
           B_{\alpha \beta \gamma \delta (K)}^{2211} & B_{\alpha \beta \gamma \delta (K)}^{2222} &
           B_{\alpha \beta \gamma \delta (K)}^{2212} & B_{\alpha \beta \gamma \delta (K)}^{2221}\\
           A_{\alpha \beta \gamma \delta (K)}^{1211} & A_{\alpha \beta \gamma \delta (K)}^{1222} &
           A_{\alpha \beta \gamma \delta (K)}^{1212} & A_{\alpha \beta \gamma \delta (K)}^{1221} &
           B_{\alpha \beta \gamma \delta (K)}^{1211} & B_{\alpha \beta \gamma \delta (K)}^{1222} &
           B_{\alpha \beta \gamma \delta (K)}^{1212} & B_{\alpha \beta \gamma \delta (K)}^{1221}\\
           A_{\alpha \beta \gamma \delta (K)}^{2111} & A_{\alpha \beta \gamma \delta (K)}^{2122} &
           A_{\alpha \beta \gamma \delta (K)}^{2112} & A_{\alpha \beta \gamma \delta (K)}^{2121} &
           B_{\alpha \beta \gamma \delta (K)}^{2111} & B_{\alpha \beta \gamma \delta (K)}^{2122} &
           B_{\alpha \beta \gamma \delta (K)}^{2112} & B_{\alpha \beta \gamma \delta (K)}^{2121} \\
             &       &       &      &      &        &           &        \\ \nonumber
           - B_{\alpha \beta \gamma \delta (K)}^{1111} & -B_{\alpha \beta \gamma \delta (K)}^{1122} &
            -B_{\alpha \beta \gamma \delta (K)}^{1112} & -B_{\alpha \beta \gamma \delta (K)}^{1121} &
           - A_{\alpha \beta \gamma \delta (K)}^{1111} & -A_{\alpha \beta \gamma \delta (K)}^{1122} &
           -A_{\alpha \beta \gamma \delta (K)}^{1112}  & -A_{\alpha \beta \gamma \delta (K)}^{1121}\\
           - B_{\alpha \beta \gamma \delta (K)}^{2211} & -B_{\alpha \beta \gamma \delta (K)}^{2222} &
           -B_{\alpha \beta \gamma \delta (K)}^{2212}  & -B_{\alpha \beta \gamma \delta (K)}^{2221} &
           - A_{\alpha \beta \gamma \delta (K)}^{2211} & -A_{\alpha \beta \gamma \delta (K)}^{2222} &
           -A_{\alpha \beta \gamma \delta (K)}^{2212}  & -A_{\alpha \beta \gamma \delta (K)}^{2221}\\
           - B_{\alpha \beta \gamma \delta (K)}^{1211} & -B_{\alpha \beta \gamma \delta (K)}^{1222} &
           -B_{\alpha \beta \gamma \delta (K)}^{1212}  & -B_{\alpha \beta \gamma \delta (K)}^{1221} &
           - A_{\alpha \beta \gamma \delta (K)}^{1211} & -A_{\alpha \beta \gamma \delta (K)}^{1222} &
           -A_{\alpha \beta \gamma \delta (K)}^{1212}  & -A_{\alpha \beta \gamma \delta (K)}^{1221} \\
          - B_{\alpha \beta \gamma \delta (K)}^{2111} & -B_{\alpha \beta \gamma \delta (K)}^{2122} &
           -B_{\alpha \beta \gamma \delta (K)}^{2112}  & -B_{\alpha \beta \gamma \delta (K)}^{2121} &
           - A_{\alpha \beta \gamma \delta (K)}^{2111} & -A_{\alpha \beta \gamma \delta (K)}^{2122} &
           -A_{\alpha \beta \gamma \delta (K)}^{2112}  & -A_{\alpha \beta \gamma \delta (K)}^{2121} \\
           \end{array} \right)\\  && \times
\left( \begin{array}{c}   {\tilde X}_{(\gamma 1 \delta 1)K}^{m}  \\ {\tilde X}_{(\gamma 2 \delta 2)K}^{m} \\
  {\tilde X}_{(\gamma 1 \delta 2)K}^{m} \\  {\tilde X}_{(\gamma 2 \delta 1)K}^{m} \\ \\
     {\tilde Y}_{(\gamma 1 \delta 1)K}^{m} \\ {\tilde Y}_{(\gamma 2 \delta 2)K}^{m} \\
     {\tilde Y}_{(\gamma 1 \delta 2)K}^{m}\\{\tilde Y}_{(\gamma 2 \delta 1)K}^{m} \end{array} \right)
 = \hbar {\Omega}_K^{m}
 \left ( \begin{array}{c} {\tilde X}_{(\alpha 1 \beta 1)K}^{m}  \\{\tilde X}_{(\alpha 2 \beta 2)K}^{m} \\
 {\tilde X}_{(\alpha 1 \beta 2)K}^{m} \\  {\tilde X}_{(\alpha 2 \beta 1)K}^{m}\\ \\
{\tilde Y}_{(\alpha 1 \beta 1)K}^{m} \\ {\tilde Y}_{(\alpha 2 \beta 2)K}^{m} \\
{\tilde Y}_{(\alpha 1 \beta 2)K}^{m} \\ {\tilde Y}_{(\alpha 2 \beta 1)K}^{m} \end{array} \right)  ~,
\end{eqnarray}
where the amplitudes
${\tilde X}^m_{(\alpha \alpha''  \beta \beta'')K }$ and ${\tilde Y}^m_{(\alpha
\alpha''  \beta \beta'')K}$ in Eq. (\ref{qrpaeq}) stand for forward and backward going amplitudes from the state ${ \alpha
\alpha'' }$ to the state  ${\beta  \beta''}$ \cite{Ha2015a} and are related to those in Eq. (\ref{phonon}) by 
$\tilde{X^m}_{(\alpha \alpha'' \beta \beta'')K}=\sqrt2 \sigma_{\alpha \alpha'' \beta \beta''} X^m_{(\alpha \alpha''
 \beta \beta'')K}$
and $\tilde{Y^m}_{(\alpha \alpha'' \beta \beta'')K}=\sqrt2 \sigma_{\alpha \alpha'' \beta \beta''}
Y^m_{(\alpha \alpha'' \beta \beta'')K}$, with a normalization constant $\sigma_{\alpha \alpha'' \beta \beta''}$ = 1 if $\alpha = \beta$ and $\alpha''$ =
$\beta''$, otherwise $\sigma_{\alpha \alpha'' \beta \beta'' }$ = $\sqrt 2$ \cite{Ch93}. 

The $A$ and $B$ matrices in Eq. (\ref{qrpaeq}) are given by
\begin{eqnarray} \label{eq:mat_A}
A_{\alpha \beta \gamma \delta (K)}^{\alpha'' \beta'' \gamma'' \delta''}  = && (E_{\alpha
   \alpha''} + E_{\beta \beta''}) \delta_{\alpha \gamma} \delta_{\alpha'' \gamma''}
   \delta_{\beta \delta} \delta_{\beta'' \delta''}
       - \sigma_{\alpha \alpha'' \beta \beta''}\sigma_{\gamma \gamma'' \delta \delta''}\\ \nonumber
   &\times&
   \sum_{\alpha' \beta' \gamma' \delta'}
   [-g_{pp} (u_{\alpha \alpha''\alpha'} u_{\beta \beta''\beta'} u_{\gamma \gamma''\gamma'} u_{\delta \delta''\delta'}
   +v_{\alpha \alpha''\alpha'} v_{\beta \beta''\beta'} v_{\gamma \gamma''\gamma'} v_{\delta \delta''\delta'} )
    ~V_{\alpha \alpha' \beta \beta',~\gamma \gamma' \delta \delta'}
    \\ \nonumber  &-& g_{ph} (u_{\alpha \alpha''\alpha'} v_{\beta \beta''\beta'}u_{\gamma \gamma''\gamma'}
     v_{\delta \delta''\delta'}
    +v_{\alpha \alpha''\alpha'} u_{\beta \beta''\beta'}v_{\gamma \gamma''\gamma'} u_{\delta \delta''\delta'})
    ~V_{\alpha \alpha' \delta \delta',~\gamma \gamma' \beta \beta'}
     \\ \nonumber  &-& g_{ph} (u_{\alpha \alpha''\alpha'} v_{\beta \beta''\beta'}v_{\gamma \gamma''\gamma'}
     u_{\delta \delta''\delta'}
     +v_{\alpha \alpha''\alpha'} u_{\beta \beta''\beta'}u_{\gamma \gamma''\gamma'} v_{\delta \delta''\delta'})
    ~V_{\alpha \alpha' \gamma \gamma',~\delta \delta' \beta \beta' }],
\end{eqnarray}
\begin{eqnarray} \label{eq:mat_B}
B_{\alpha \beta \gamma \delta (K)}^{\alpha'' \beta'' \gamma'' \delta''}  =
 &-& \sigma_{\alpha \alpha'' \beta \beta''} \sigma_{\gamma \gamma'' \delta \delta''}
  \\ \nonumber &\times&
 \sum_{\alpha' \beta' \gamma' \delta'}
  [g_{pp}
  (u_{\alpha \alpha''\alpha'} u_{\beta \beta''\beta'}v_{\gamma \gamma''\gamma'} v_{\delta \delta''\delta'}
   +v_{\alpha \alpha''\alpha'} v_{{\bar\beta} \beta''\beta'}u_{\gamma \gamma''\gamma'} u_{{\bar\delta} \delta''\delta'} )
   ~ V_{\alpha \alpha' \beta \beta',~\gamma \gamma' \delta \delta'}\\ \nonumber
     &- & g_{ph} (u_{\alpha \alpha''\alpha'} v_{\beta \beta''\beta'}v_{\gamma \gamma''\gamma'}
     u_{\delta \delta''\delta'}
    +v_{\alpha \alpha''\alpha'} u_{\beta \beta''\beta'}u_{\gamma \gamma''\gamma'} v_{\delta \delta''\delta'})
   ~ V_{\alpha \alpha' \delta \delta',~\gamma \gamma' \beta \beta'}
     \\ \nonumber  &- & g_{ph} (u_{\alpha \alpha''\alpha'} v_{\beta \beta''\beta'}u_{\gamma \gamma''\gamma'}
      v_{\delta \delta''\delta'}
     +v_{\alpha \alpha''\alpha'} u_{\beta \beta''\beta'}v_{\gamma \gamma''\gamma'} u_{\delta \delta''\delta'})
   ~ V_{\alpha \alpha' \gamma \gamma',~\delta \delta' \beta \beta'}],
\end{eqnarray}
where the $u_{\alpha \alpha''\alpha'}$ and $v_{\alpha \alpha''\alpha'}$ coefficients are determined from the Hartree-Fock-Bogoliubov transformation between particles and quasi-particles with isospin $\alpha'$ and $\alpha''$, respectively, in the $\alpha$ state \cite{Ha18-1}. The $g_{\text{pp}}$ and $g_{\text{ph}}$ stand for particle-particle and particle-hole renormalization factors for the residual interactions in Eqs. (\ref{eq:mat_A}) and (\ref{eq:mat_B}).  The two-body interactions $V_{\alpha \beta,~\gamma \delta}$ and $V_{\alpha \delta,~\gamma \beta}$ are particle-particle and particle-hole matrix elements of the residual $N$-$N$ interaction $V$, respectively, which are calculated from the $G$-matrix {as solutions of the Bethe-Goldstone equation from the CD-Bonn potential.}  

{The two-body interactions $V_{\alpha \beta,~\gamma \delta}$ and $V_{\alpha \delta,~\gamma \beta}$ correspond to the $p-p$ and $p-h$ channels, respectively, of the residual $N-N$ interaction in the deformed state. They are 
calculated from the $G$-matrix in the spherical basis as follows
\begin{eqnarray}
V_{\alpha \alpha' \beta \beta' ,~\gamma \gamma' \delta \delta'} 
= - && \sum_{J} \sum_{abcd} F_{\alpha a {\bar\beta} b}^{JK}  F_{\gamma c {\bar \delta} d}^{JK} G( a \alpha' b \beta' c \gamma' d \delta' , J)~, \\ \nonumber
V_{\alpha \alpha' \delta \delta' , \gamma \gamma' \beta \beta'} 
= && \sum_J \sum_{abcd} F_{\alpha a {\delta} d}^{JK}  F_{\gamma
 c {\beta} b}^{JK} G( a \alpha' d \delta' c \gamma' b \beta' , J)~, \\ \nonumber
V_{\alpha \alpha' \gamma \gamma' , \delta \delta' \beta \beta'} 
 = && \sum_J \sum_{abcd} F_{\alpha a {\gamma} c}^{JK}  F_{\beta
 b {\delta} d}^{JK} G( a \alpha' c \gamma' d \delta' b \beta' , J)~.
\end{eqnarray} 
Here $a$ and $\alpha$ indicates spherical and deformed SPS, respectively. {We note that the isospin $\alpha', \beta', \gamma'$ and $\delta'$ in the $G$-matrix is {replaced} by total isospin ($T=0$ or $T=1$) of the two-body interaction in the isospin representation.} We use $F_{\alpha a {\bar \beta} b}^{JK} = B_{a}^{\alpha} B_b^{\beta} C_{j_a \Omega_\alpha j_b \Omega_\beta}$ for $K = \Omega_{\alpha} + \Omega_{\beta}$. The expansion coefficient $B_{\alpha}^a$ is defined as 
\begin{equation}
| \alpha \Omega_{\alpha} > =\sum_{a} B_{a}^{\alpha} |a \Omega_{\alpha} > ~,~  B_{a}^{\alpha} = \sum_{N n_z} C_{l \Lambda { 1 \over 2} \Sigma}^{j \Omega_\alpha} A_{N n_z \Lambda}^{ N_0 l} b_{N n_z \Sigma},  
\end{equation}
with the Clebsch-Gordan coefficient $C_{l \Lambda { 1 \over 2} \Sigma}^{j \Omega_\alpha}$, the spatial overlap integral $A_{N n_z \Lambda}^{ N_0 l}$, and the eigenvalues obtained from the total Hamiltonian in the deformed basis $b_{N n_z \Sigma}$. Detailed formulas regarding the transformation of Eq.~(7) can be found in Ref. \cite{Ha2015a}.}

The GT transition operator ${\hat {\textrm{O}} }_{1\mu}^{\pm}$ is defined by
 \begin{equation} \label{eq:btop}
{\hat {\textrm{O}} }_{1 \mu}^{-}  = \sum_{\alpha \beta }
< \alpha p |\tau^{+} \sigma_K | \beta n >  c_{\alpha p}^{\dagger} {c}_{\beta n},~
{\hat {\textrm{O}}}_{1 \mu}^{+}  =  {( { \hat T}_{1 \mu}^{-} )}^\dagger  = {(-)}^{\mu}
{ \hat {\textrm{O}}}_{1,-\mu}^{-}.
\end{equation}
Detailed calculation for the transformation from the intrinsic frame to the nuclear laboratory system were presented in Ref. \cite{Ha2015a}. 
The ${\hat {\textrm{O}}}^{\pm}$ transition amplitudes from the ground state of an initial (parent) nucleus
to the excited state of a final (daughter) nucleus, {\it i.e.} the one phonon state
$\vert K^+, m\rangle$, are written as
\begin{eqnarray} \label{eq:phonon}
&&< K^+, m | {\hat {\textrm{O}}}_{K }^- | ~QRPA >  \\ \nonumber
&&= \sum_{\alpha \alpha''\rho_{\alpha} \beta \beta''\rho_{\beta}}{\cal N}_{\alpha \alpha''\rho_{\alpha}
 \beta \beta''\rho_{\beta} }
 < \alpha \alpha''p \rho_{\alpha}|  \sigma_K | \beta \beta''n \rho_{\beta}>
 [ u_{\alpha \alpha'' p} v_{\beta \beta'' n} X_{(\alpha \alpha''\beta \beta'')K}^{m} +
v_{\alpha \alpha'' p} u_{\beta \beta'' n} Y_{(\alpha \alpha'' \beta \beta'')K}^{m}], \\ \nonumber
&&< K^+, m | {\hat {\textrm{O}}}_{K }^+ | ~QRPA >  \\ \nonumber
&&= \sum_{\alpha \alpha'' \rho_{\alpha} \beta \beta''\rho_{\beta}}{\cal N}_{\alpha \alpha'' \beta \beta'' }
 < \alpha \alpha''p \rho_{\alpha}|  \sigma_K | \beta \beta''n \rho_{\beta}>
 [ u_{\alpha \alpha'' p} v_{\beta \beta'' n} Y_{(\alpha \alpha'' \beta \beta'')K}^{m} +
v_{\alpha \alpha'' p} u_{\beta \beta'' n} X_{(\alpha \alpha'' \beta \beta'')K}^{m} ]~,
\end{eqnarray}
where $|~QRPA >$ denotes the correlated QRPA ground state in the intrinsic frame and
the nomalization factor is given as $ {\cal N}_{\alpha \alpha'' \beta
 \beta''} (J) = \sqrt{ 1 - \delta_{\alpha \beta} \delta_{\alpha'' \beta''} (-1)^{J + T} }/
 (1 + \delta_{\alpha \beta} \delta_{\alpha'' \beta''}).$ 

Here, {for the BCS calculation,} we use the following %energy density functional for the 
pairing interaction commonly adopted in the SHF approach, 
\be \label{eq:pairing}
{\cal H}_{pair} ({\bf r}) = {1 \over 2} V_0 [ 1 - V_1 ({\rho / \rho_0  })^{\gamma} ], 
\ee
where the parameter $V_0$ {is adjusted by fixing the GT peak position in $^{18}$O,} and the other {parameters} $V_1$ and $\gamma$ are taken from Ref. \cite{Stoitsov}. The {values} of the parameters and the pairing gaps  obtained with the pairing window {associated with an energy} cut-off of 60 MeV, are summarized in Table \ref{tab:beta2} for {the Skyrme EDF parameter sets used in the present work. The reduced spin-orbit force is 0.6 of the original strength \cite{Sagawa04}.}   

We adopt the pairing strengths listed in Table I for the following reason. {Since $^{12}$C exhibits sub-magic behavior in both neutron and proton sectors, we determine the pairing strengths by referring instead to the open-shell nucleus $^{16}$C. With the standard and reduced spin-orbit interaction, the adopted coupling yields reasonable and finite neutron pairing gaps, as shown in Fig.~\ref{fig:16C pairings}.} 
{We note that in the present DQRPA formalism the $T=0$ pairing is not considered. The $T=0$ pairing might be strong in $N\approx Z$ nuclei due to the maximal overlap of the spatial part of wave function, and may give a substantial effect on GT transition strenght. It remains  for future work to include both $T=0$ and $T=1$ pairings in the present formalism and to apply for  nuclei in the vicinity of $N= Z$ line in the nuclear chart.}

\begin{figure} %Fig.1
\includegraphics[width=0.8\linewidth]{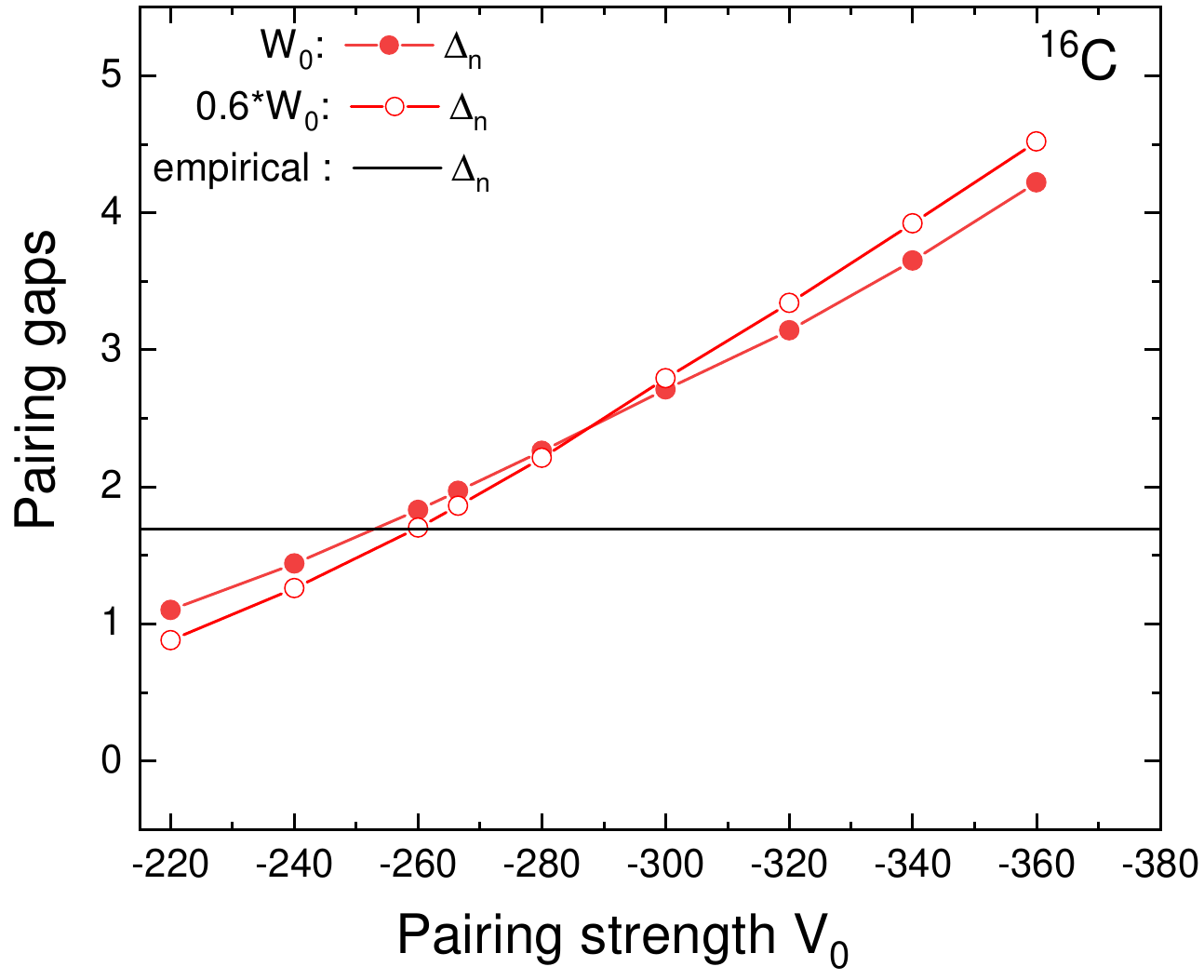}
\caption{{(Color online) Neutron pairing gaps of $^{16}$C vs the pairing strength, {when the mean-field is
obtained using the SGII set, for standard and reduced spin-orbit strength ($W_0$ and $0.6 \times W_0$). They are compared with  the empirical pairing gaps obtained by {the five-point formula}.} Relevant potential energy curves for each case are presented, respectively, in Figs.~2 and 7. 
}}
\label{fig:16C pairings}
\end{figure}

\begin{table} % Table 1
\caption[bb]{The parameters relevant to the pairing interaction in the present calculation together with the deformation parameter $\beta_2$ obtained from the PEC by DSHFB in Fig.~\ref{fig:pec}. The value in the parenthesis for $^{12}$C is the deformation {obtained through} the reduction of the spin-orbit strength.}
\setlength{\tabcolsep}{2.0 mm}
\begin{tabular}{ccccccc}\hline
                                            & & & &  & $\beta_2$   \\
Skyrme  EDF & $V_0$ & {$V_1$} & {$\gamma$}  &[12C,  &14C,         & 16C]  \\ \hline \hline
 SLy4    & -295.3  & 0.5 & 1.0    & [0.0, &  0.0, &  0.04]~~~~~~~~~    \\
 SkP      & -220.2 & 0.5 & 1.0    & [0.0,  & 0.0,  &  0.00]~~~~~~~~~    \\
 SGII     & -266.5  & 0.5 & 1.0   & [0.0 ($-$0.34),  &  0.0,  &  0.14]  \\ \hline
 \end{tabular}
\label{tab:beta2}
\end{table}
\begin{figure} % Fig.2
\includegraphics[width=1.0\linewidth]{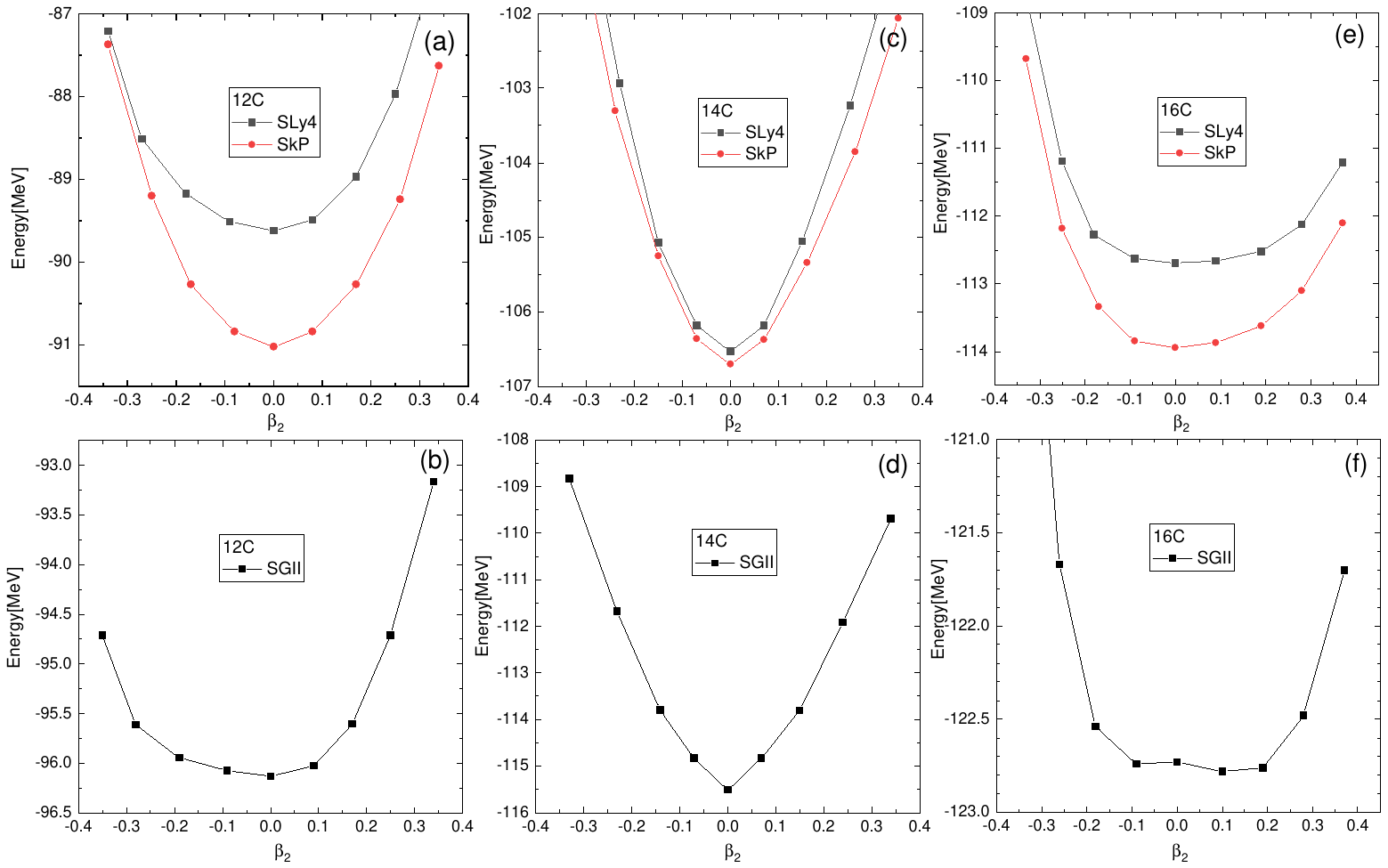}
\caption{(Color online)  Potential energy curves (PECs) of $^{12, 14,16}$C with SLy4, Skp, and SGII, respectively. 
}
\label{fig:pec}
\end{figure}
\section{Results and discussions}
\begin{figure} %Fig.3
\includegraphics[width=0.55\linewidth]{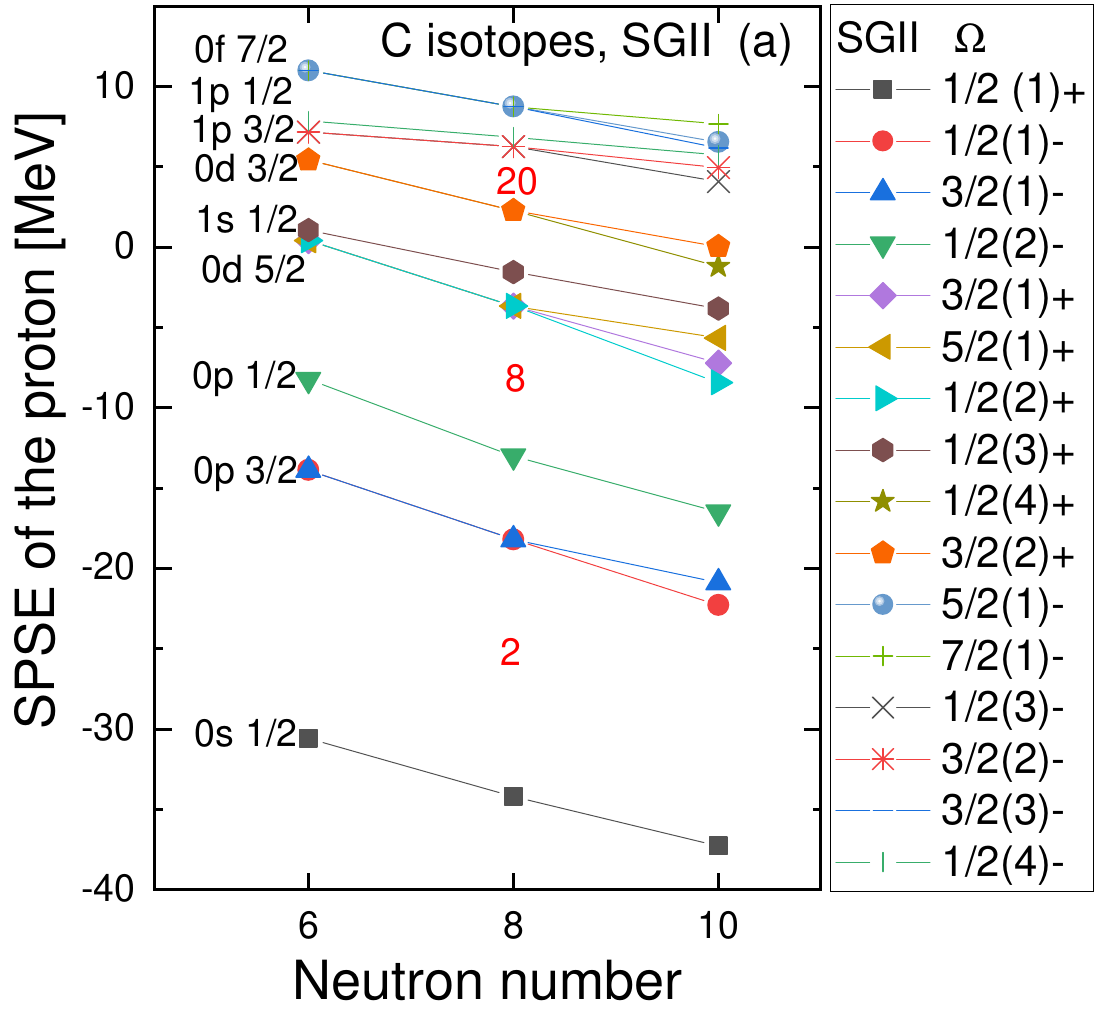}
\includegraphics[width=0.41\linewidth]{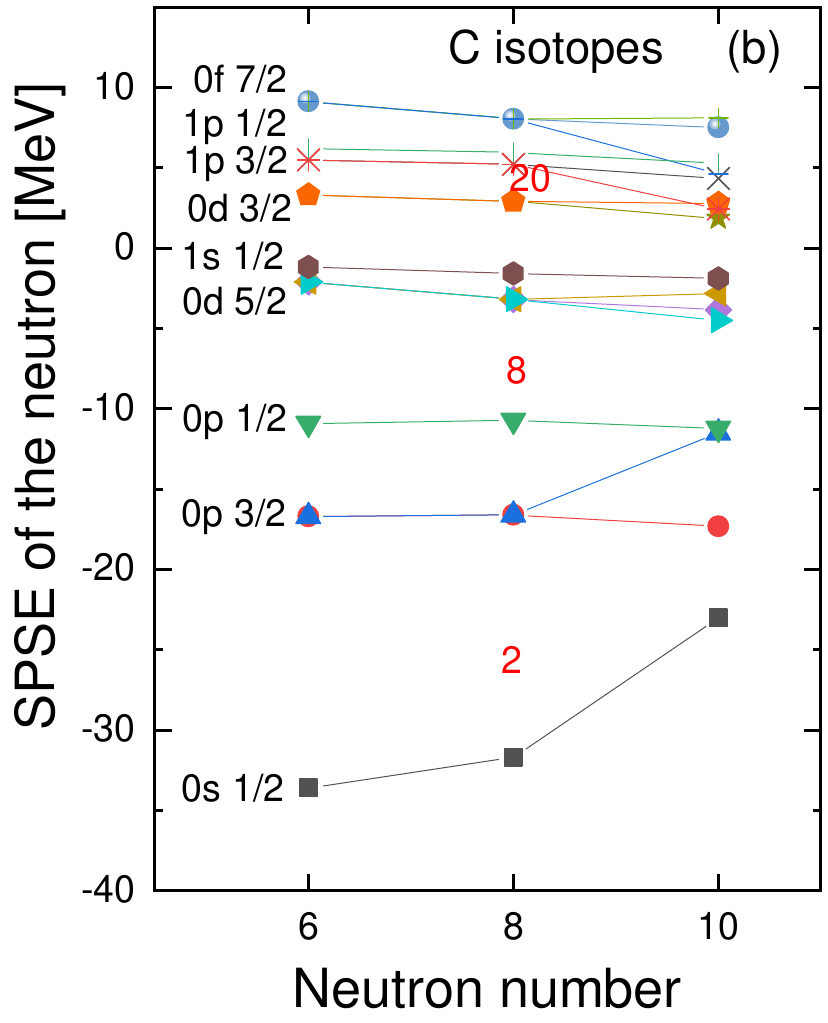}
\caption{(Color online) SPSE (single particle state energy) of protons (a) and neutrons (b) of carbon isotopes with SGII.}
\label{fig:c_sps}
\end{figure}
This study is based on the mean field obtained with the axially deformed Skyrme Hartree-Fock-Bogoliubov (DSHFB) approach using a harmonic oscillator basis \cite{Stoitsov}. The particle model space, for the nuclei considered here, {includes states} up to $N$ = 18 $\hbar \omega$  for the deformed and spherical basis. The pairing energy cut is 60 MeV with mixed pairing and SkP, SLy4, and SGII. In the DQRPA, we adopt the wave functions and the SPS energies from the DSHFB equations.  
We utilize the SPS energies from the deformed SHF mean field, and the $G$-matrix for the two-body interaction in Eq. (\ref{eq:mat_A}) and (\ref{eq:mat_B}) is also calculated by using the wave functions  from the deformed SHF mean field.
As shown in Fig.~\ref{fig:pec}, $^{12}$C and $^{14}$C nuclei {show spherical shapes} with SkP and SLy4. In contrast, $^{16}$C is spherical with SkP and SLy4, but $\beta_{2}$ is 0.14 with SGII, as summarized in Table \ref{tab:beta2}. 
In Fig.~\ref{fig:c_sps}, we illustrate the SPS evolution of protons and neutrons with the deformation. We note that the SPS energy splitting of either the $p_{3/2}$ state or the $d_{5/2}$ state in $^{16}$C, 
induced by deformation, reduces  the shell gaps, and consequently the neutron energy gap
 between $3/2_{1}^{-}$ from  $0p_{3/2}$ and $1/2_{2}^{-}$ 
from $0p_{1/2}$
 {state disappears}. 
 
%%%%%%%%%%%%%%%%%%%% 12C %%%%%%%%%%%%%%%%%%%%%%%%%%%%%%%%%%%%%%%%%%%%%%%
\subsection{GT strength distributions for $^{12}$C}

\begin{figure} %Fig.4
\includegraphics[width=0.45\linewidth]{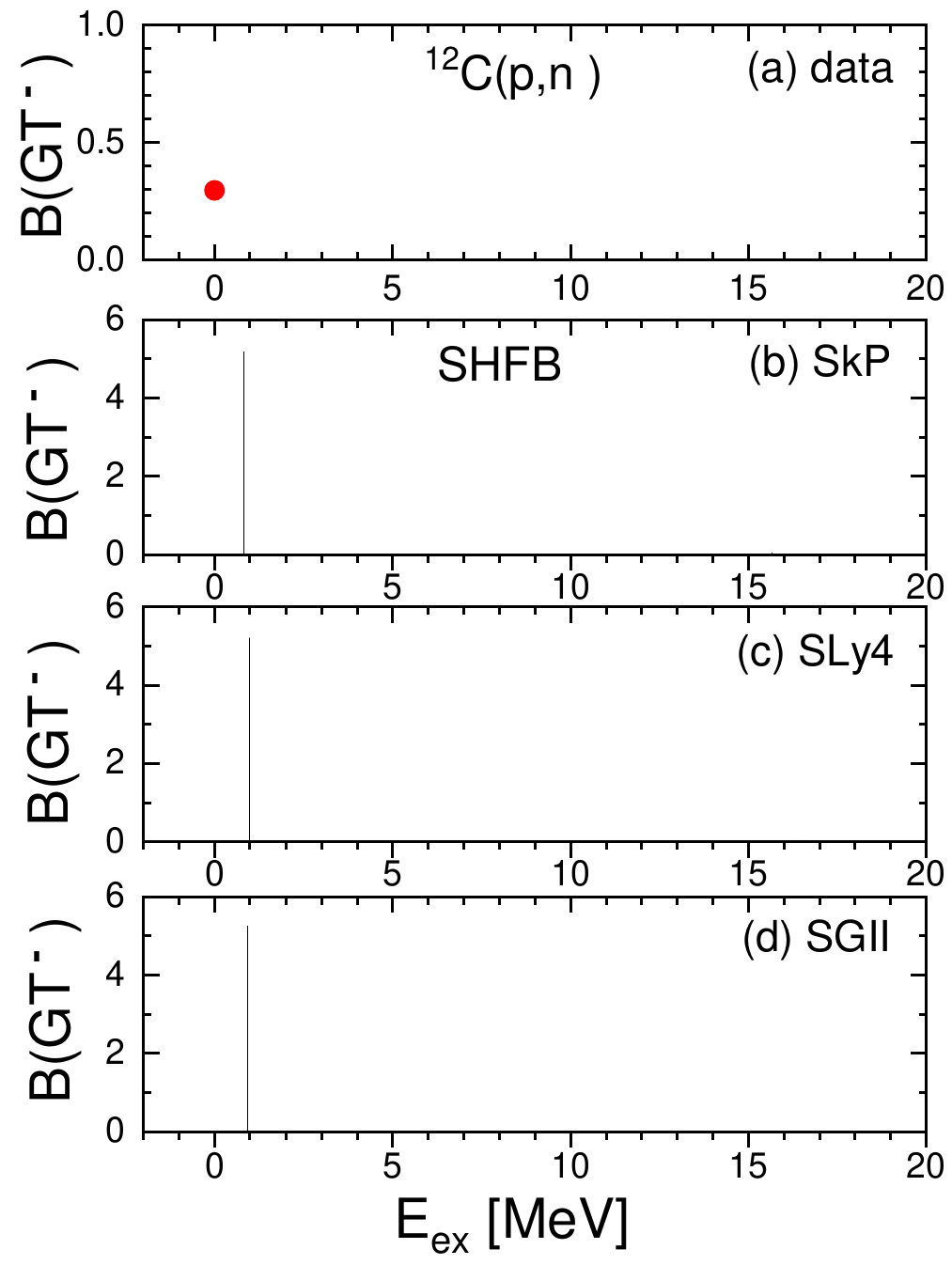}
\includegraphics[width=0.45\linewidth]{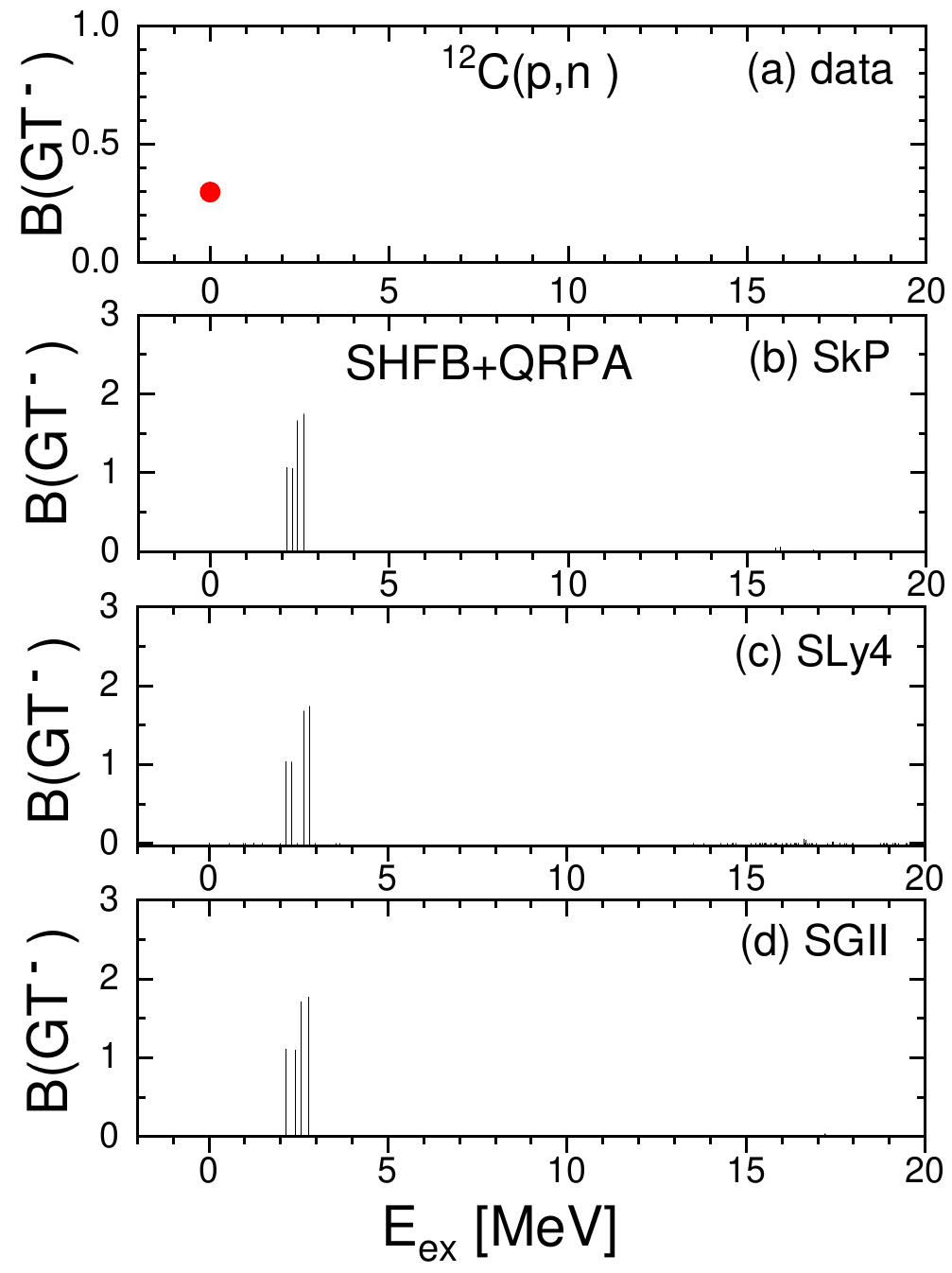}
\caption{(Color online) The B(GT$^{(-)}$) transition strength distributions from $^{12}$C with SkP, SLy4, and SGII are displayed, respectively. Experimental data are taken from Ref. \cite{Anderson96}.}
\label{fig:12c_gtm2}
\end{figure}

\begin{figure} % fig.5
\includegraphics[width=0.65\linewidth]{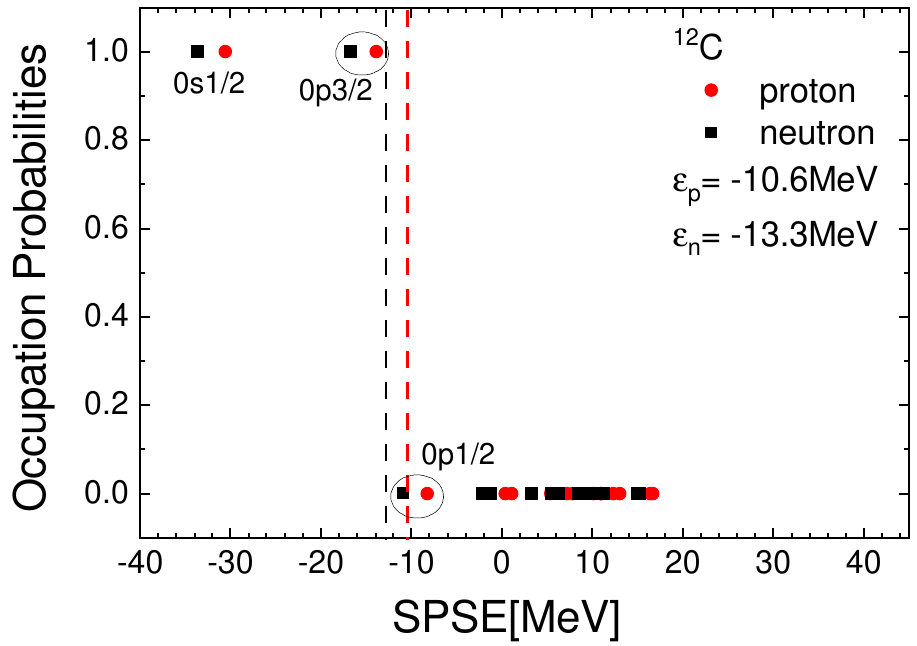}
\caption{(Color online) Occupation probabilities of protons {(red circles) and neutrons (black squares)} in $^{12}$C. The red (black) dashed line is the Fermi energy of protons (neutrons). Occupation probabilities of SPSs of $0p_{3/2}$ and $0p_{3/2}$ of proton and neutron around Fermi energy are not smeared by the pairing interaction. }
\label{fig:12c_occu}
\end{figure}

In Fig.~\ref{fig:pec}, the ground state of $^{12}$C is shown to be spherical for SkP, SLy4, and SGII. We present the GT$^{(-)}$ strength,  
B(GT$^{(-)}$), for $^{12}$C in Fig.~\ref{fig:12c_gtm2}. Low-energy GT$^{(-)}$ strength may significantly influence $\beta$-decay and electron capture rates in the stellar environments. It is interesting to see that GT$^{(-)}$ strengths are concentrated in the low-energy region both in experiment and theory. However, the calculated energy peak deviates from the experimental value by about 3 MeV. This feature is almost irrespective of the Skyrme parameter sets. {The residual interaction
by the QRPA shifts the single DSHFB GT peak by about a few MeV upward and fragments it as shown by the difference between the left and right panels.}

\begin{table} % Table II
\caption[bb]{Excited energy, B(GT), main configurations, and forward amplitudes for the GT$^{(-)}$ state of $^{12}$C around $\approx 2-3$ MeV in the panel (c) of Fig.~\ref{fig:12c_gpp_g2}. {We note that the K = 0 (1) part is mainly affected by the $p-p$ ($p-h$) interaction.}\\
 }
\setlength{\tabcolsep}{2.0 mm}
\begin{tabular}{ccccc}\hline
       K            & $E_{ex}$     &  B(GT)     &  configuration (spherical limit)         & { $X^{2}$ }       \\ \hline \hline
       0            &  2.15            &  1.10        &$ \pi 0p_{3/2}$, $\nu 0p_{1/2}$       & 0.50   \\
                     &                   &               &{$ \pi 0p_{1/2}$, $\nu 0p_{3/2}$}      & 0.49   \\  
                     &  2.40           &  1.09       &$ \pi 0p_{3/2}$, $\nu 0p_{1/2}$       & 0.50   \\ 
                     &                   &              &{$ \pi 0p_{1/2}$, $\nu 0p_{3/2}$}       & 0.49   \\\hline
       1            &  2.58            & 1.70       &$ \pi 0p_{3/2}$, $\nu 0p_{1/2}$        & 0.59   \\
                    &                    &              &{$ \pi 0p_{1/2}$, $\nu 0p_{3/2}$}       & 0.39   \\
                    &  2.78             &  1.76       &$ \pi 0p_{3/2}$, $\nu 0p_{1/2}$      & 0.63   \\
                    &                    &              &{$ \pi 0p_{1/2}$, $\nu 0p_{3/2}$}       & 0.33   \\\hline  \hline
 \end{tabular}
\label{tab:12config}
\end{table}

\begin{figure} % fig 6
\includegraphics[width=0.65\linewidth]{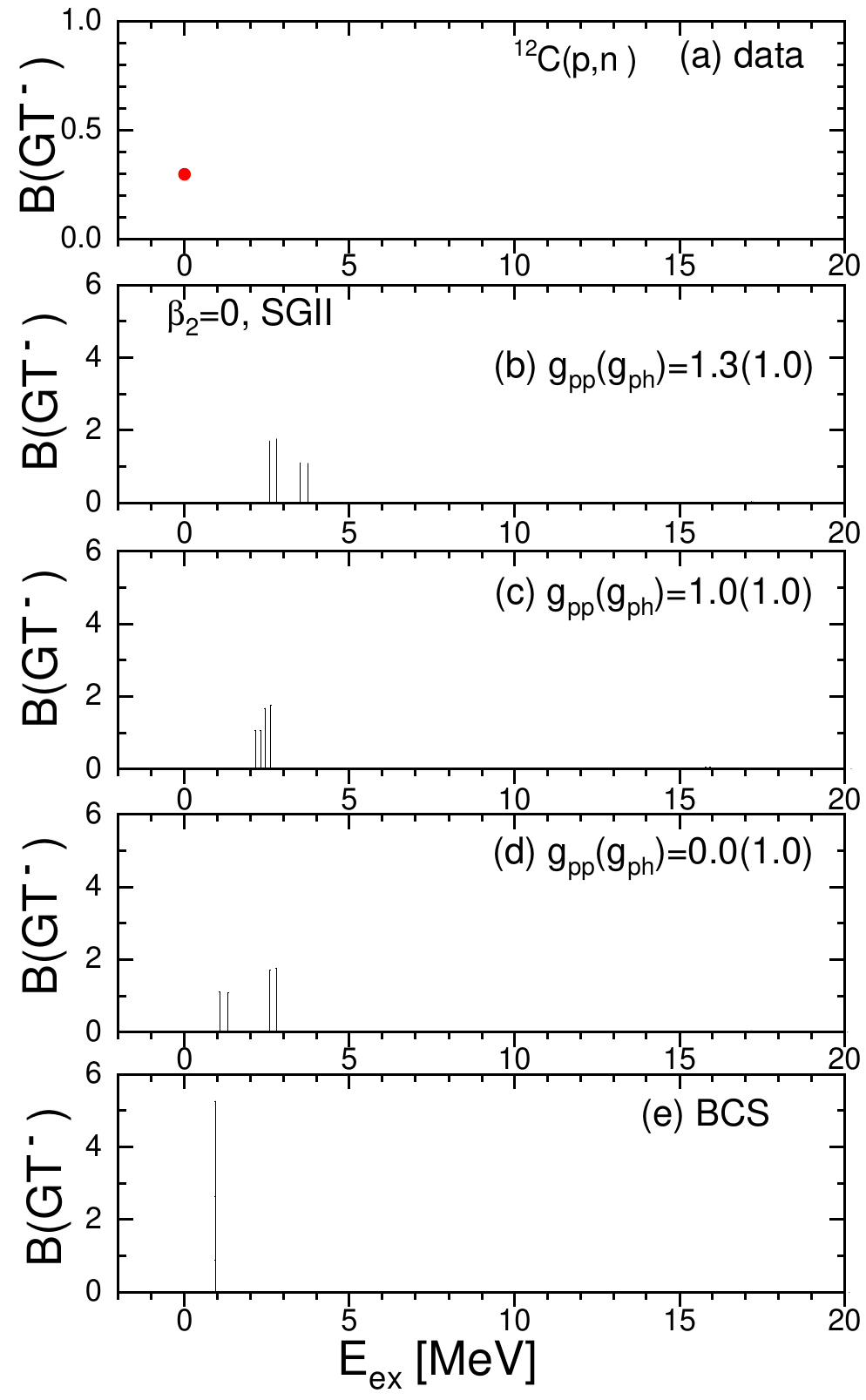}
\caption{(Color online) {Effect of the $p-p$ interaction on the GT$^{(-)}$ strength distributions in $^{12}$C, calculated with the SGII interaction, and compared with the data taken from \cite{Anderson96}. Panels} (b), (c), and (d) show the results where the normalization factor $g_{\text{ph}}$=1.0 is fixed in all the calculations, but the factor $g_{\text{pp}}$ is changed as 0.0, 1.0, and 1.3. Panel (e) is calculated without RPA.
}
\label{fig:12c_gpp_g2}
\end{figure}
The major configuration of the low-energy peak of the GT$^{(-)}$ {strength} turns out to be the {$\nu {0p3/2} \rightarrow \pi {0p1/2}$ configuration channel by $p-h$} excitation as detailed in Table \ref{tab:12config}. {We note that the $\nu {0p1/2} \rightarrow \pi {0p3/2}$ configuration channel by $p-h$ excitation showing almost the same magnitude $|X|^2$ does not contribute to GT$^{(-)}$ transition, but only contribute to the GT$^{(+)}$ transition by the occupation probabilities.}
The neutron in the $\nu 0p3/2$ state cannot go the same proton orbit because both proton and neutron $0p3/2$ states are fully occupied, as shown in Fig.~\ref{fig:12c_occu}: the pairing interaction does not produce noticeable smearing of occupation probabilities around the Fermi surface.

%%%%%%%%%%%%%%%%%% g_pp and g_hh effects

Then, we examine the role played by the $p-h$ and $p-p$ interactions in the GT strength distribution of $^{12}$C in Fig.~\ref{fig:12c_gpp_g2}. {A clear effect  of the repulsive $p-p$ interaction on the GT strength  can be seen in Figs.~\ref{fig:12c_gpp_g2} (b)-(d).} {The two small peaks around 1 MeV in the panel (d) shift to $E_{ex} \simeq $ 3.5 MeV with QRPA correlations, but the  shift is opposite to the one needed to reproduce the experimental observation. The $p-h$ interaction does not shift the GT strengths. It is interesting that the pattern is different in spherical
heavy nuclei, where $p-h$ interaction is dominant and $p-p$ interaction is attractive albeit weak.} {This may come from a different 
interplay of these interactions in deformed QRPA. More careful discussions regarding the role of the $p-p$ and $p-h$ interactions in light deformed nuclei are necessary for further understanding.}

\begin{figure} %Fig.7
\includegraphics[width=1.0\linewidth]{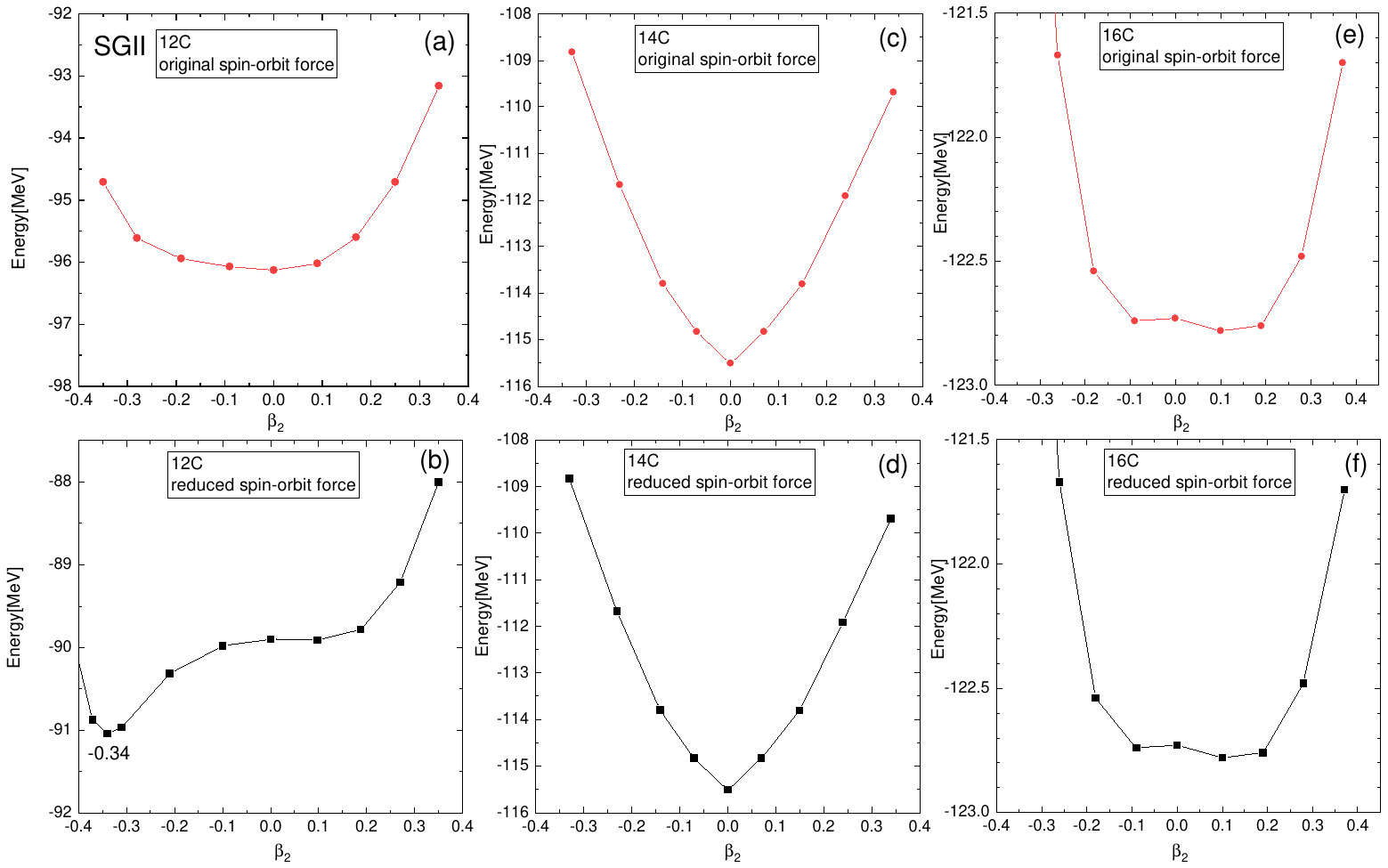}
\caption{(Color online)  {{PECs of $^{12, 14,16}$C with SGII using either the reduced spin-orbit force (b,d,f) or the original one (a,b,c), respectively.}}  
}
\label{fig:pec_reso}
\end{figure}

\begin{figure} %Fig.8
\includegraphics[width=0.55\linewidth]{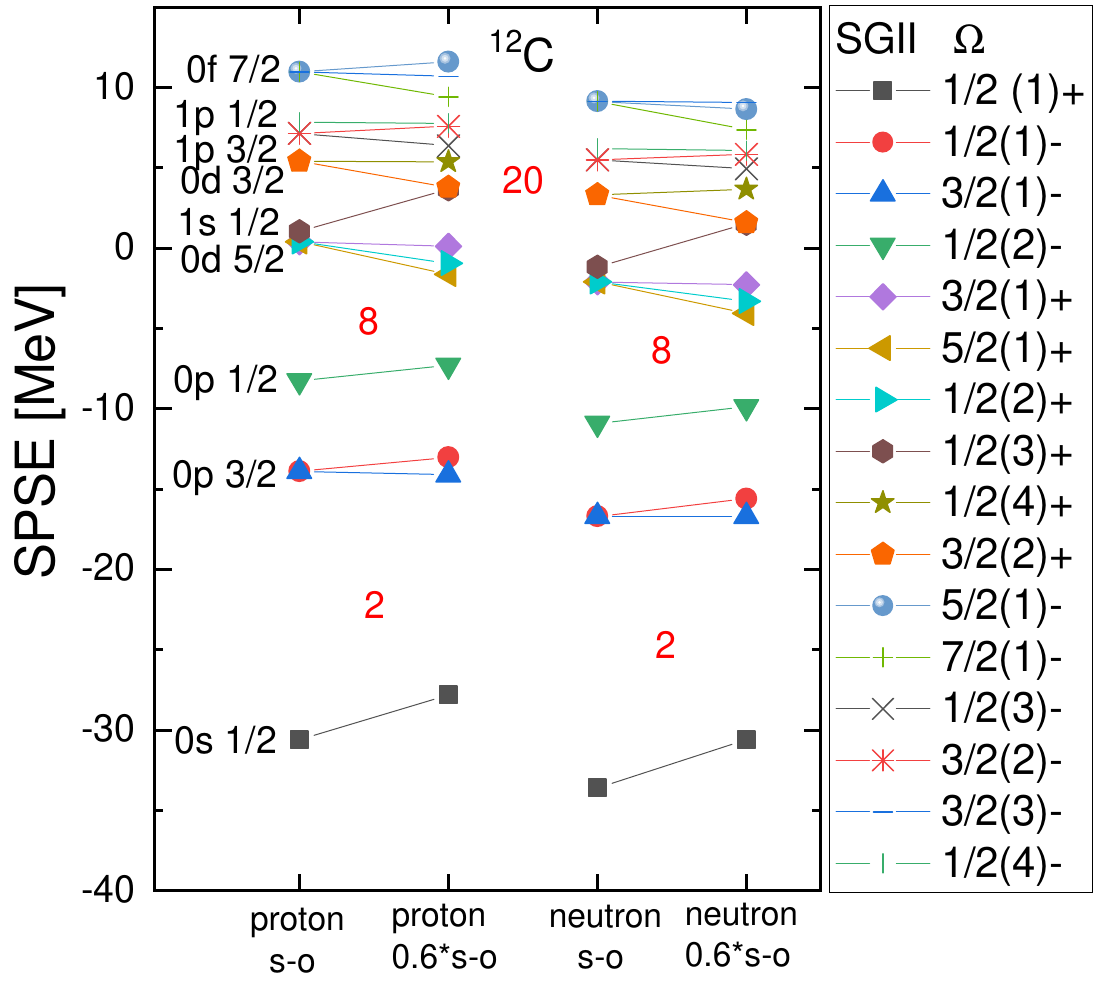}
\caption{(Color online) Change of the SPSE (single particle state energy) of protons (a) and neutrons (b) by the reduction of the  spin-orbit coupling strength in $^{12}$C with SGII.}
\label{fig:c_sps_reso}
\end{figure}

\begin{figure} Fig. 9
\includegraphics[width=0.45\linewidth]{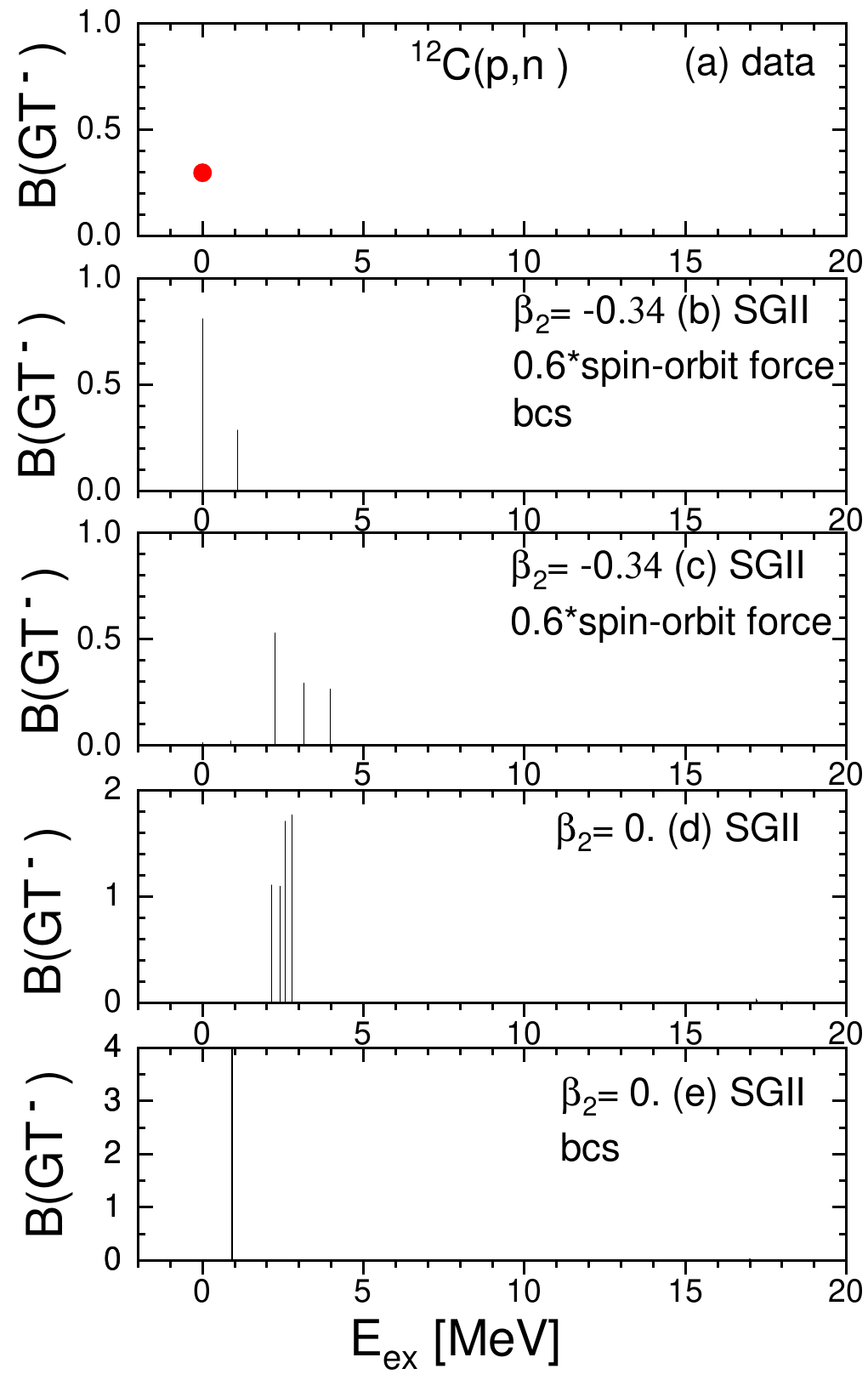}
\caption{(Color online) The B(GT$^{(-)}$) of $^{12}$C with SGII using both the reduced spin-orbit force and the original one are displayed, respectively.}
\label{fig:12c_g2_reso}
\end{figure}
%
%

%%%%%%%%%%%%%%%%%%%%%%%%%%%%%%

%%%%%%%%%%%%%%%%%% Spin-orbit force effects

In Fig.~\ref{fig:pec_reso}, we present the PE curves obtained with the reduced spin-orbit interaction in SGII, which predict an oblate deformation with $\beta_2$= -0.34 for $^{12}$C, but only indiscernible change in $^{14,16}$C. {The changes in the SPS energies, induced by the reduction of the spin-orbit, are presented in Fig.~\ref{fig:c_sps_reso}.} The deformation effect due to the reduction of the spin-orbit strength appears clearly in the Nilsson basis.

The B(GT) results obtained by using the reduced spin-orbit force are shown  in  Fig.~\ref{fig:12c_g2_reso}: there is a peak around E = 0 MeV region in the BCS case, which is apparently compatible with the experimental data. But, similarly to the results {that we have already shown in Figs. 4 and 6, the residual interaction also shifts the strength upward.} 

{The choice of reducing the spin-orbit strength to 60 \% of its original value follows Ref.~\cite{Sagawa04}, where such a reduction was shown to improve the description of binding energies and shell evolution in light p-shell nuclei. In the present case, the weakened spin-orbit splitting enhances configuration mixing between the 0$p_{3/2}$ and 0$p_{1/2}$ orbitals, leading to a GT peak closer to the experimental position.}

{In brief, the low-lying GT state peaks around $E_{ex}$ = 0 MeV is explained at the BCS stage, but the residual interaction shifts the GT peak to a few MeV higher. This trend is almost irrespective of the Skyrme parameter sets. The reduction of the spin-orbit strength also succeeds in explaining the GT peak around $E_{ex}$ = 0 MeV {in the BCS case,} but the residual interaction shifts again the peak a few MeV higher, similarly to the results by the original Skyrme spin-orbit interaction strength.}

%
%%%%%%%%%%%%%%%%%%%%% 14C %%%%%%%%%%%%%%%%%%%%%%%%%%%%%%%%%%%%%%%%%%%%%%%
%
%
%
\subsection{$^{14}$C}

\begin{figure} %Fig.10
\includegraphics[width=0.65\linewidth]{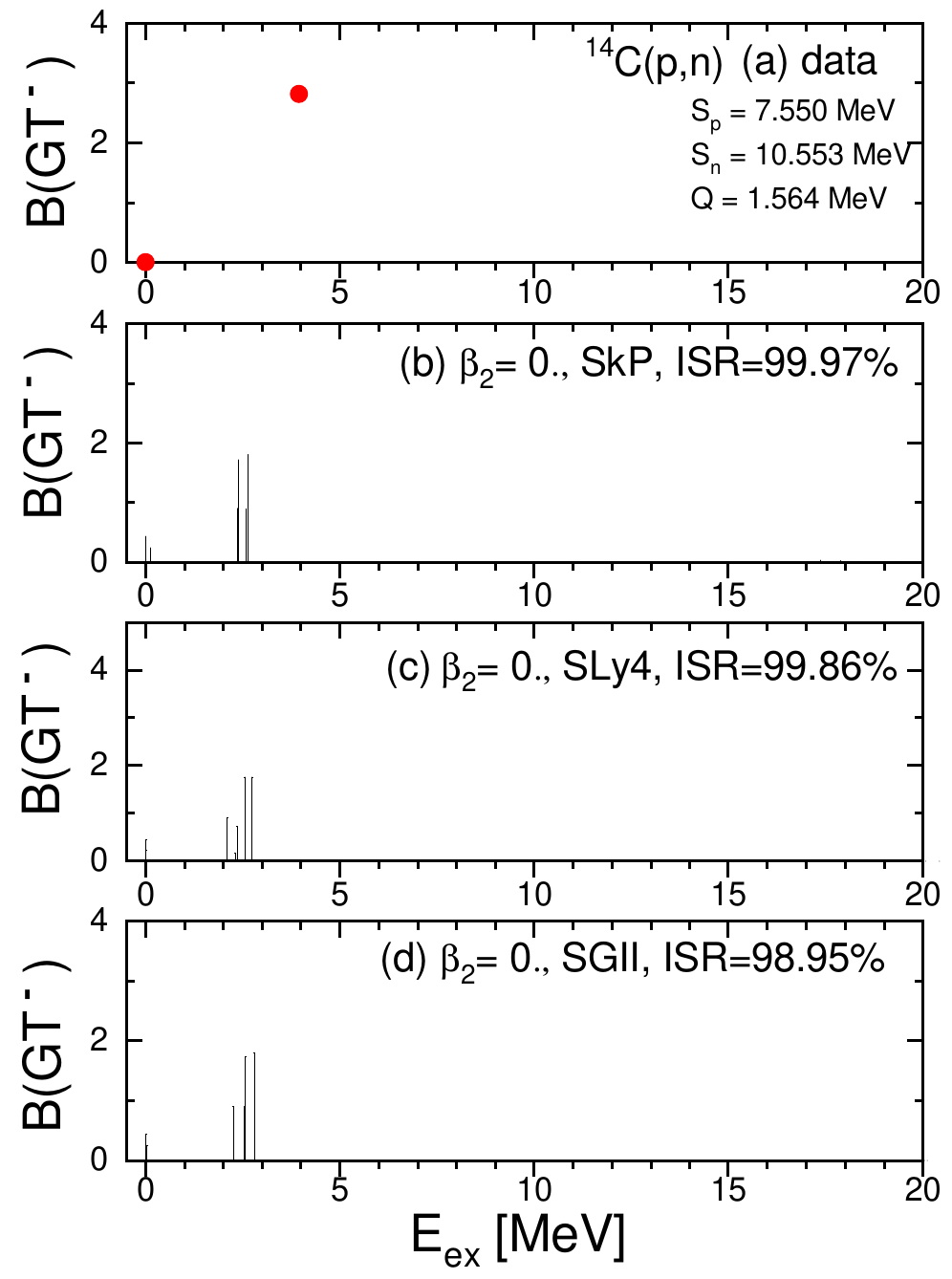}
\caption{(Color online) The B(GT$^{(-)}$) transition strength distributions in $^{14}$C with SkP, SLy4, and SGII are displayed, respectively. Experimental data are taken from Ref. \cite{Anderson91}.
 }
\label{fig:14c_gtm}
\end{figure}
Hereafter, we discuss the results for $^{14}$C, which shows two GT peaks, {$\approx$ 0 MeV and $\approx$ 2--3 MeV} with three Skyrme parameter sets as shown in Fig.~\ref{fig:14c_gtm}. {A unique feature of GT strength in $^{14}$C is the strong GT$^{(-)}$ strength in the excited $1^+$ state at 4 MeV above the ground $1^+$ state, while $^{12}$C and $^{6}$He show strong GT strength in the 1$^+$ ground state ($^6$Li) or the first excited $1^+$ state near the ground state of the daughter nucleus \cite{Ayala2022}.} This is because the {$\nu0p1/2\rightarrow \pi0p1/2$} channel is the main $p$-$h$ configuration for the lowest 1$^+$ state for $^{14}$C and the excited state has the {$\nu0p3/2\rightarrow \pi0p1/2$} channel as the main configuration: the latter configuration has 8 times larger $B(GT:0^+\rightarrow 1^+)$ strength than the former. 
On the other hand, in $^{6}$He, the lower GT configuration channel is {$\nu0p3/2\rightarrow \pi0p3/2$} and
the higher one is {$\nu0p3/2\rightarrow \pi0p1/2$}, and the lower energy $p-h$ state has 25\% larger B(GT) strength than the latter.
In another picture describing  the same physical mechanism, the GT strength distribution  of $^{14}$N  can be attributed to an 
intrinsic pair of proton and neutron hole configuration of the doubly magic $^{16}$O core as discussed in Ref. \cite{Fujita2020}.

In the present case, the GT strength concentrates in the second $1^{+}$ state, while the lowest $1^{+}$ is weakly excited. 
Our calculations reproduce the two peaks, but the position of the second peak lies a bit below the experimental value.
Figure \ref{fig:14c_gpp} shows the effects of residual interaction, $p-p$ interaction and $p-h$ interaction, on the B(GT$^{(-)}$) strength distribution in $^{14}$C. {In the panel (d) of Fig.~\ref{fig:14c_gpp}}, the increase of the strengths of $g_{pp}$ and $g_{ph}$ by about 30~\%  shifts the second peak (denoted by the green dashed line), corresponding to the summed B(GT) strength in the 3-4 MeV region, toward higher excitation energies, and provides a good description of the experimental data. 
The main configurations for the peaks at $E_{ex} \approx 0-4$ MeV  in Fig.~\ref{fig:14c_gpp} (d) are presented in Table \ref{tab:14config}. We note that the $\nu 0 p_{1/2} \rightarrow \pi 0 p_{1/2}$ transition  also contributes to the B(GT) value on top of the main $\nu 0 p_{3/2} \rightarrow \pi 0 p_{1/2}$ one, which is dominant in the case of $^{12}$C.

\begin{figure} %Fig.11
\includegraphics[width=0.65\linewidth]{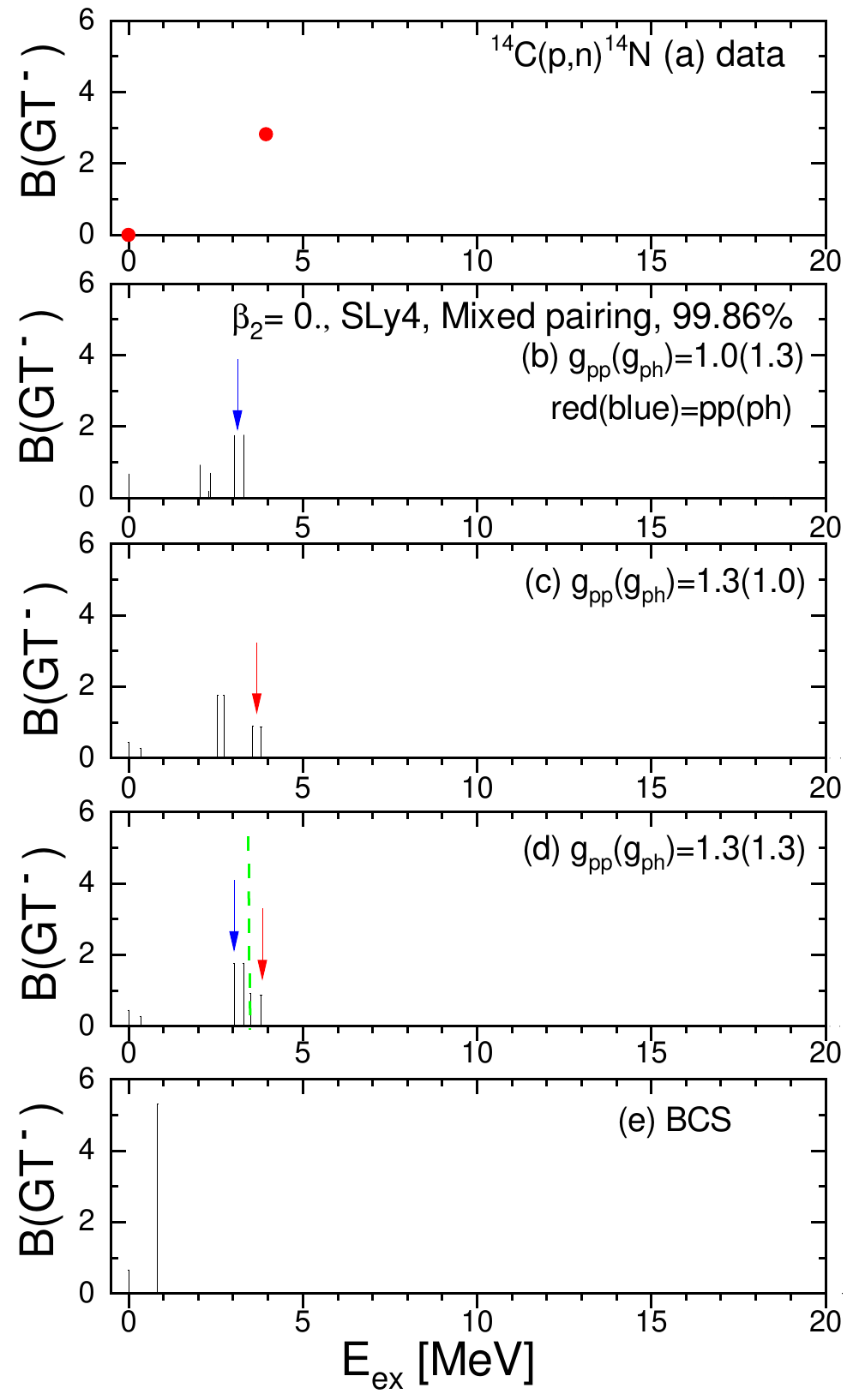}
\caption{(Color online) Effect of the $p-p$ (red) and $p-h$ (blue) interaction on the GT$^{(-)}$ transition strength distributions in $^{14}$C. In the results of the panels (b), (c), and (d), {the parameters $g_{\text{pp}}$ and $g_{\text{ph}}$} are changed from 1.0 to 1.3. The green dashed line represents the summed B(GT) values in the 3-4 MeV region. Panel (e) shows the results with BCS. Here, the red (blue) arrows indicate the effect of the $p-p$ ($p-h$) interaction.
}
\label{fig:14c_gpp}
\end{figure}

\begin{table}  % Table III
\caption[bb]{Excitation energy, B(GT), main configurations, and forward amplitudes for the GT$^{(-)}$ states in $^{14}$C between $\approx 0-4$ MeV in the panel (d) of Fig.~\ref{fig:14c_gpp}. \\
 }
\setlength{\tabcolsep}{2.0 mm}
\begin{tabular}{ccccc}\hline
       K            & $E_{ex}$     &  B(GT)     &  configuration (spherical limit)         & { $X^{2}$ }         \\ \hline \hline
       0            &  0.36            &  0.28       &$ \pi 0p_{3/2}$, $\nu 0p_{1/2}$       & 0.50   \\
                     &                   &               &{$ \pi 0p_{1/2}$, $\nu 0p_{3/2}$}      & 0.49   \\  
                     &  3.50            &  0.91       &$ \pi 0p_{3/2}$, $\nu 0p_{1/2}$       & 0.50   \\
                     &                   &               &{$ \pi 0p_{1/2}$, $\nu 0p_{3/2}$}      & 0.49   \\ 
                     &  3.80            &  0.87      &$ \pi 0p_{3/2}$, $\nu 0p_{1/2}$       & 0.50   \\
                     &                   &               &{$ \pi 0p_{1/2}$, $\nu 0p_{3/2}$}      & 0.49   \\\hline                    
       1            &  0.00            &  0.44       &$ \pi 0p_{1/2}$, $\nu 0p_{1/2}$        & 1.00   \\
                     &  3.03            &  1.74       &$ \pi 0p_{3/2}$, $\nu 0p_{1/2}$       & 0.59   \\
                     &                   &               &{$ \pi 0p_{1/2}$, $\nu 0p_{3/2}$}      & 0.39   \\ 
                     &  3.30            &  1.76       &$ \pi 0p_{3/2}$, $\nu 0p_{1/2}$         & 0.63   \\
                     &                   &               &{$ \pi 0p_{1/2}$, $\nu 0p_{3/2}$}      & 0.33   \\\hline 

 \end{tabular}
\label{tab:14config}
\end{table}

%%%%%%%%%%%%%%%%%%%% 16C %%%%%%%%%%%%%%%%%%%%%%%%%%%%%%%%%%%%%%%%%%%%%%%
\subsection{$^{16}$C}

Finally, we discuss the results for $^{16}$C, which is prolate deformed with SGII. Therefore we consider the DSHFB solution associated with $\beta_2 = 0.14$ in the following.  We expect the GT strength distribution of $^{16}$C to exhibit both low-lying GT peaks, similar to those in $^{14}$C, and high-lying GT peaks, because of two extra neutrons on top of the magic number $N=8$. The occupation probabilities of the neutron in $0p_{1/2}$ and $0d_{5/2}$ orbitals near the Fermi energy are smeared due to the deformation and the open-shell structure, as shown in Fig.~\ref{fig:16c_occu}, which may lead to the generation of the high-lying GT states above $15$ MeV. Indeed, Fig.~\ref{fig:16c_gtm} shows  such results for the B(GT$^{(-)}$) strength distribution of $^{16}$C.

The results show low-lying peaks as well as GT strength  above 15 MeV. The main configurations for the peaks below $E_{ex} \approx 5.0$ MeV and above $E_{ex} \approx 15$ MeV  in Fig.~\ref{fig:16c_gtm}(c) are presented in Table \ref{tab:16config}. The  peaks below $5.0$ MeV come from the configuration $\nu 0p_{1/2} \rightarrow \pi 0p_{3/2}$. However, the $0.19$ MeV peak is attributed to the $\nu 0p_{1/2} \rightarrow \pi 0p_{1/2}$ {configuration}. In contrast, the peaks above $15$ MeV come from the configuration $\nu 0d_{5/2} \rightarrow \pi 0d_{5/2}$ and 
$\nu 0d_{3/2} \rightarrow \pi 0d_{5/2}$. As shown in panel (d) in Fig.~\ref{fig:16c_gtm}, in the case of SGII, the peaks are more strongly fragmented in high-lying GT states due to deformation compared with the SkP and SLy4 parameter sets in the panel (a) and (b) in Fig.~\ref{fig:16c_gtm}.

{The reduction of the Ikeda sum rule (ISR) exhaustion in the deformed SGII case (approximately 88 \%) reflects both deformation-induced configuration mixing and the finite model space employed in the present calculation. In particular, the fragmentation of strength into high-lying states and possible contributions from continuum configurations, especially the $\nu0d_{5/2}\rightarrow \pi0d3/2$ channel,  may lead to a partial quenching of the summed GT strength within the adopted model space. The prolate deformation predicted by SGII ($\beta_2 \sim 0.14$) reduces the effective shell gap at $N$ = 8 through the splitting of Nilsson orbitals derived from the 0$d_{5/2}$ configuration. This reduction facilitates additional neutron excitations into higher-lying states, thereby generating the high-lying GT strength observed above 15 MeV.}

\begin{figure} %Fig.12
\includegraphics[width=0.65\linewidth]{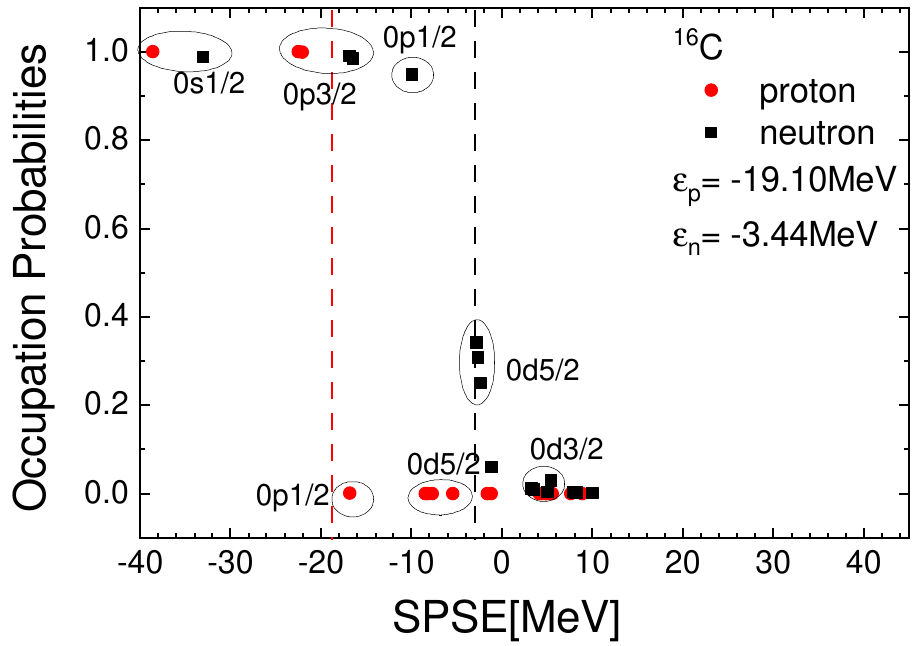}
\caption{(Color online) Occupation probabilities of protons and neutrons in $^{16}$C. The red (black) dashed line is the Fermi energy of protons (neutrons). Occupation probabilities of {the neutron SPSs $0p_{1/2}$ and $0d_{5/2}$} around the Fermi energy are smeared due to the deformation and the open-shell structure.}
\label{fig:16c_occu}
\end{figure}
\begin{figure} %Fig.13
\includegraphics[width=0.65\linewidth]{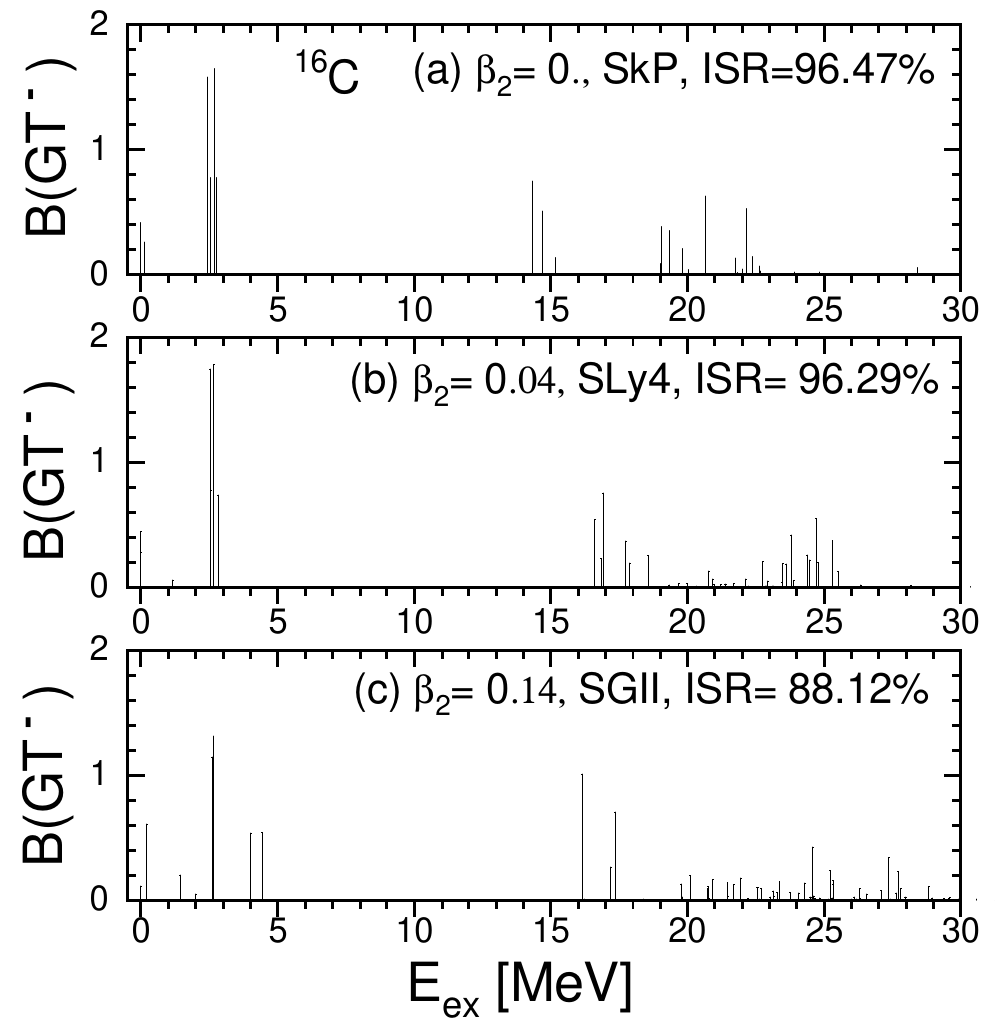}
\caption{(Color online)  The B(GT$^{(-)}$) transition strength distributions in $^{16}$C with SkP, SLy4, and SGII are displayed, respectively. }
\label{fig:16c_gtm}
\end{figure}
\begin{table} % Table IV
\caption[bb]{Excited energy, B(GT), main configurations, and forward amplitudes for the GT$^{-}$ state of $^{16}$C  at $E_{ex} \approx 0-30$ MeV in the panel (c) of Fig.~\ref{fig:16c_gtm}. \\
 }
\setlength{\tabcolsep}{2.0 mm}
\begin{tabular}{ccccc}\hline
       K            & $E_{ex}$      &  B(GT)     &  configuration (spherical limit)         & { $X^{2}$ }        \\ \hline \hline
       0            &  0.19            &  0.60       &$ \pi 0p_{1/2}$, $\nu 0p_{1/2}$       & 0.78   \\
                     &                    &               &$ \pi 0d_{5/2}$, $\nu 0d_{5/2}$       & 0.15   \\
                     &  4.00            &  0.53       &$ \pi 0p_{3/2}$, $\nu 0p_{1/2}$       & 0.50   \\
                     &                   &               &{$ \pi 0p_{1/2}$, $\nu 0p_{3/2}$}      & 0.49   \\ 
                     &  4.42            &  0.58       &$ \pi 0p_{3/2}$, $\nu 0p_{1/2}$       & 0.50   \\ 
                     &                   &               &{$ \pi 0p_{1/2}$, $\nu 0p_{3/2}$}      & 0.49   \\ \hline                   
       1            &  2.58            &  1.14       &$ \pi 0p_{3/2}$, $\nu 0p_{1/2}$        & 0.56   \\
                     &                   &               &{$ \pi 0p_{1/2}$, $\nu 0p_{3/2}$}      & 0.41   \\
                     &  2.65            &  1.31       &$ \pi 0p_{3/2}$, $\nu 0p_{1/2}$       & 0.51   \\
                      &                   &               &{$ \pi 0p_{1/2}$, $\nu 0p_{3/2}$}      & 0.48   \\
                     &  16.14          &  1.00       &$ \pi 0d_{5/2}$, $\nu 0d_{5/2}$       & 1.00   \\
                     &  17.35          &  1.76       &$ \pi 0d_{5/2}$, $\nu 0d_{5/2}$       & 0.98   \\
                     &  24.57          &  0.42       &$ \pi 0d_{3/2}$, $\nu 0d_{5/2}$       & 0.49   \\
                     &                   &               &$ \pi 0d_{5/2}$, $\nu 0d_{3/2}$       & 0.49   \\
                     &  27.37          &  0.33       &$ \pi 0d_{3/2}$, $\nu 0d_{5/2}$       & 0.49   \\
                     &                   &               &$ \pi 0d_{5/2}$, $\nu 0d_{3/2}$       & 0.49   \\\hline \hline
 \end{tabular}
\label{tab:16config}
\end{table}

\begin{table} % Table V
\caption[bb]{Excited energy, B(GT$^{(-)}$), main configurations, and forward amplitudes for the GT$^{(-)}$ states of $^{16}$C at $E_{ex} \approx 0-30$ MeV in the panel (c) of Fig.~\ref{fig:16c_gtm}. \\
 }
\setlength{\tabcolsep}{2.0 mm}
\begin{tabular}{ccccc}\hline
       K            & $E_{ex}$      &  B(GT)     &  configuration (spherical limit)         & X        \\ \hline \hline
       0            &  0.19            &  0.60       &$ \pi 0p_{1/2}$, $\nu 0p_{1/2}$       & 0.77   \\
                     &                    &               &$ \pi 0d_{5/2}$, $\nu 0d_{5/2}$       & 0.05   \\
                     &  4.00            &  0.53       &$ \pi 0p_{3/2}$, $\nu 0p_{1/2}$       & 0.99   \\
                     &  4.42            &  0.58       &$ \pi 0p_{3/2}$, $\nu 0p_{1/2}$       & 0.99   \\\hline                    
       1            &  2.58            &  1.14       &$ \pi 0p_{3/2}$, $\nu 0p_{1/2}$        & 0.97   \\
                     &  2.65            &  1.31       &$ \pi 0p_{3/2}$, $\nu 0p_{1/2}$       & 0.98   \\
                     &  16.14          &  1.00       &$ \pi 0d_{5/2}$, $\nu 0d_{5/2}$       & 1.00   \\
                     &  17.35          &  1.76       &$ \pi 0d_{5/2}$, $\nu 0d_{5/2}$       & 0.70   \\
                     &  24.57          &  0.42       &$ \pi 0d_{3/2}$, $\nu 0d_{5/2}$       & 0.70   \\
                     &  27.37          &  0.33       &$ \pi 0d_{3/2}$, $\nu 0d_{5/2}$       & 0.70   \\\hline \hline
 \end{tabular}
\label{tab:16config}
\end{table}

\section{Summary and Discussions}

We have investigated the GT strength distributions of carbon isotopes, $^{12,14,16}$C, in the
framework of Deformed QRPA (DQRPA). The present calculation adopts the DSHF mean
field for the calculation of the single-particle states, whereas the DQRPA has the residual
interaction derived from a $G$-matrix calculation: the residual forces in the DQRPA are
computed from the realistic CD-Bonn potential, and introduced independently from the
Skyrme force used to generate the mean field. The underlying idea behind the use of the
residual interaction from the CD-Bonn potential is to include the finite-range pairing interactions
by the $G$-matrix. The purpose of the work is also to use a more realistic Skyrme-type mean-field
than the deformed Woods Saxon (DWS) potential {that we have used in previous works; fully self-consistent QRPA using the Skyrme type interaction as a residual force is left for a future work.}

{In $^{12}$C, the unperturbed $p-h$ excitation spectrum is characterised by a low-lying GT peak near zero excitation energy. However, the inclusion of residual correlations shifts this peak upward by approximately 3 MeV. A reduction of the spin-orbit interaction improves the description of the experimental B(GT) strength, indicating that spin-orbit effects and deformation play an essential role in light p-shell nuclei.}

In $^{14}$C, we confirm that the second GT peak of $^{14}$C shift to higher energy by the effect of $p-p$ and $p-h$ interactions, which enables a good reproduction of the two GT peaks observed in the experimental data.

For $^{16}$C, deformation leads to additional fragmentation and the appearance of high-lying GT strength above 15 MeV, originating mainly from 
0$d_{5/2}$ and 0$d_{3/2}$ configurations. {Although experimental data of GT transitions for 16C are not yet available, we have a robust theoretical prediction.}

{The present results demonstrate that deformation and residual correlations must be treated consistently in order to describe GT transitions in light nuclei. In particular, the sensitivity of $^{12}$C and $^{16}$C to spin-orbit strength and deformation indicates that light p-shell nuclei provide a stringent test of nuclear energy density functionals in the spin-isospin channel but not only.}

Finally, we emphasize that the present formalism does not include $T=0$ pairing correlations. Since $T=0$ pairing may play an important role in nuclei near the $N = Z$ line, its inclusion constitutes an important direction for future work. In light N = Z nuclei such as $^{12}$C, $T = 0$ (isoscalar) pairing correlations may compete with the conventional $T = 1$ pairing channel and significantly modify the low-lying GT strength. The inclusion of $T = 0$ pairing is therefore expected to affect both the collectivity and the energy position of the lowest GT states. A systematic study including both pairing channels will be addressed in a forthcoming work.

Although the present approach combines a Skyrme mean field with a realistic $G$-matrix residual interaction and is therefore not fully self-consistent, this hybrid framework allows us to incorporate finite-range correlations beyond standard Skyrme-based QRPA calculations. A fully self-consistent treatment using the same energy density functional in both mean field and residual channels will be investigated in future works.

\section*{Acknowledgement}
This work was supported by the National Research Foundation of Korea (Grant Nos. RS-2025-00513410 and RS-2024-00460031).  The work of MKC is supported by the National Research Foundation of Korea (Grant Nos. RS-2021-NR060129 and RS-2025-16071941). 
The work of HS is supported by Chinese Academy of
Sciences (CAS) President’s International Fellowship Initiative
(PIFI) Grant No. 2024PVA0003-Y1.

\end{document}